\documentstyle[amsmath,amscd]{article}
\def\be{\begin{eqnarray}}
\def\ee{\end{eqnarray}}
\def\nn{\nonumber}

\newcommand{\e}{\epsilon}
\newcommand{\al}{\alpha}
\newcommand{\Up}{\Upsilon}
\newcommand{\tT}{\tilde T}
\newcommand{\Pm}{{\cal P}_-}
\newcommand{\Pp}{{\cal P}_+}
\newcommand{\mPm}{\left(-{\cal P}_-\right)}
\newcommand{\lnM}{\log{\frac{M^2}{\mu^2}}}
\newcommand{\go}{\begin{picture}(16,10)\put(0,3){\line(1,0){14}}
\put(7,3){\circle*{3}}\end{picture}}
\newcommand{\goo}{\begin{picture}(20,10)\put(0,3){\line(1,0){20}}
\put(5,3){\circle*{3}}\put(15,3){\circle*{3}}\end{picture}}
\newcommand{\gooo}{\begin{picture}(32,10) \put(0,3){\line(1,0){30}}
\put(5,3){\circle*{3}}\put(15,3){\circle*{3}}\put(25,3){\circle*{3}}
\end{picture}}
\newcommand{\goooo}{\begin{picture}(42,10) \put(0,3){\line(1,0){40}}
\put(5,3){\circle*{3}}\put(15,2){\circle*{3}}
\put(25,3){\circle*{3}}\put(35,2){\circle*{3}}
\end{picture}}
\newcommand{\ibox}{\begin{picture}(20,10) \put(0,3){\line(1,0){20}}
\put(10,2){\circle*{3}}
\put(4,-2){\line(0,1){8}} \put(4,6){\line(1,0){12}}
\put(16,-2){\line(0,1){8}}\put(4,-2){\line(1,0){12}}
\end{picture}}
\newcommand{\iibox}{\begin{picture}(25,10) \put(0,3){\line(1,0){20}}
\put(5,2){\circle*{3}}\put(15,2){\circle*{3}}
\put(3,-2){\line(0,1){8}} \put(3,6){\line(1,0){14}}
\put(17,-2){\line(0,1){8}}\put(3,-2){\line(1,0){14}}
\end{picture}}
\newcommand{\iiiboxleft}{\begin{picture}(38,10)
\put(-3,3){\line(1,0){36}}
\put(5,2){\circle*{3}}\put(15,2){\circle*{3}}\put(25,2){\circle*{3}}
\put(3,-2){\line(0,1){8}} \put(3,6){\line(1,0){18}}
\put(21,-2){\line(0,1){8}}\put(3,-2){\line(1,0){18}}
\end{picture}}
\newcommand{\iiiboxright}{\begin{picture}(38,10)
\put(-3,3){\line(1,0){36}}
\put(5,2){\circle*{3}}\put(15,2){\circle*{3}}\put(25,2){\circle*{3}}
\put(9,-2){\line(0,1){8}} \put(9,6){\line(1,0){18}}
\put(27,-2){\line(0,1){8}}\put(9,-2){\line(1,0){18}}
\end{picture}}
\newcommand{\iiiboxleftnet}{\begin{picture}(38,14)
\put(-3,3){\line(1,0){36}}
\put(5,2){\circle*{3}}\put(15,2){\circle*{3}}
\put(25,2){\circle*{3}}
\put(3,-2){\line(0,1){8}} \put(3,6){\line(1,0){18}}
\put(21,-2){\line(0,1){8}}\put(3,-2){\line(1,0){18}}
\put(0,-4){\line(0,1){12}} \put(0,8){\line(1,0){30}}
\put(30,-4){\line(0,1){12}}\put(0,-4){\line(1,0){30}}
\end{picture}}
\newcommand{\iiiboxrightnet}{\begin{picture}(38,14)
\put(-3,3){\line(1,0){36}}
\put(5,2){\circle*{3}}\put(15,2){\circle*{3}}
\put(25,2){\circle*{3}}
\put(9,-2){\line(0,1){8}} \put(9,6){\line(1,0){18}}
\put(27,-2){\line(0,1){8}}\put(9,-2){\line(1,0){18}}
\put(0,-4){\line(0,1){12}} \put(0,8){\line(1,0){30}}
\put(30,-4){\line(0,1){12}}\put(0,-4){\line(1,0){30}}
\end{picture}}
\newcommand{\lo}{\begin{picture}(10,10)
\put(5,3){\circle*{3}}\end{picture}}
\newcommand{\loo}{\begin{picture}(20,10)\put(10,3){\circle{10}}
\put(15,3){\circle*{3}}\end{picture}}
\newcommand{\looo}{\begin{picture}(20,10)\put(10,3){\circle{10}}
\put(5,3){\circle*{3}}\put(15,3){\circle*{3}}\end{picture}}
\newcommand{\loooo}{\begin{picture}(20,10)\put(10,3){\circle{10}}
\put(6,1){\circle*{3}}\put(15,1){\circle*{3}}
\put(10,8){\circle*{3}}\end{picture}}
\newcommand{\rtro}{\begin{picture}(20,10)
\put(0,0){\line(1,0){6}} \put(3,0){\line(0,1){10}}
\put(3,10){\circle*{2}} \put(3,10){\line(0,1){10}}
\put(3,10){\line(-3,2){10}}\put(3,10){\line(3,2){10}}
\end{picture}}
\newcommand{\rtrou}{\begin{picture}(20,30)
\put(0,0){\line(1,0){6}} \put(3,0){\line(0,1){10}}
\put(3,10){\circle*{2}} \put(3,10){\line(0,1){10}}
\put(3,10){\line(-3,2){10}}\put(3,10){\line(3,2){10}}
\put(-8,17){\circle*{2}} \put(-8,17){\line(0,1){10}}
\put(-8,17){\line(-3,2){10}}\put(-8,17){\line(3,2){10}}
\end{picture}}
\newcommand{\rtrod}{\begin{picture}(20,20)
\put(0,0){\line(1,0){6}} \put(3,0){\line(0,1){10}}
\put(3,10){\circle*{2}} \put(3,10){\line(0,1){10}}
\put(3,10){\line(-3,2){10}}\put(3,10){\line(3,2){10}}
\put(3,20){\circle*{2}} \put(3,20){\line(0,1){10}}
\put(3,20){\line(-3,2){10}}\put(3,20){\line(3,2){10}}
\end{picture}}
\newcommand{\rtrom}{\begin{picture}(20,20)
\put(0,0){\line(1,0){6}} \put(3,0){\line(0,1){10}}
\put(3,10){\circle*{2}} \put(3,10){\line(0,1){10}}
\put(3,10){\line(-3,2){10}}\put(3,10){\line(3,2){10}}
\put(13,17){\circle*{2}} \put(13,17){\line(0,1){10}}
\put(13,17){\line(-3,2){10}}\put(13,17){\line(3,2){10}}
\end{picture}}

\newcommand{\tro}{\begin{picture}(30,15)\put(0,3){\line(1,0){14}}
\put(14,3){\circle*{3}}\put(14,3){\line(1,1){10}}
\put(14,3){\line(1,-1){10}}\end{picture}}
\newcommand{\troo}{\begin{picture}(35,26)\put(0,3){\line(1,0){14}}
\put(14,3){\circle*{3}}\put(14,3){\line(1,1){20}}
\put(14,3){\line(1,-1){20}} \put(26,-8){\line(0,1){22}}
\put(26,-8){\circle*{3}}\put(26,14){\circle*{3}}
\end{picture}}
\newcommand{\trood}{\begin{picture}(35,26)\put(0,3){\line(1,0){14}}
\put(14,3){\circle*{3}}\put(14,3){\line(1,1){20}}
\put(14,3){\line(1,-1){20}} \put(26,-8){\line(0,1){22}}
\put(26,-8){\circle*{3}}\put(26,14){\circle*{3}}
\put(26,-2){\line(-1,0){7}}\put(26,-2){\circle*{2}}
\put(19,-2){\circle*{2}}
\end{picture}}
\newcommand{\troou}{\begin{picture}(35,26)\put(0,3){\line(1,0){14}}
\put(14,3){\circle*{3}}\put(14,3){\line(1,1){20}}
\put(14,3){\line(1,-1){20}} \put(26,-8){\line(0,1){22}}
\put(26,-8){\circle*{3}}\put(26,14){\circle*{3}}
\put(26,8){\line(-1,0){7}}\put(26,8){\circle*{2}}
\put(19,8){\circle*{2}}
\end{picture}}
\newcommand{\troom}{\begin{picture}(35,26)
\put(0,3){\line(1,0){14}}
\put(14,3){\circle*{3}}\put(14,3){\line(1,1){20}}
\put(14,3){\line(1,-1){20}} \put(26,-8){\line(0,1){22}}
\put(26,-8){\circle*{3}}\put(26,14){\circle*{3}}
\put(18,0){\line(0,1){7}}\put(18,-1){\circle*{2}}
\put(18,7){\circle*{2}}
\end{picture}}
\newcommand{\trooud}{\begin{picture}(35,26)
\put(0,3){\line(1,0){14}}
\put(14,3){\circle*{3}}\put(14,3){\line(1,1){20}}
\put(14,3){\line(1,-1){20}} \put(26,-8){\line(0,1){22}}
\put(26,-8){\circle*{3}}\put(26,14){\circle*{3}}
\put(26,8){\line(-1,0){7}}\put(26,8){\circle*{2}}
\put(19,8){\circle*{2}}
\put(26,-2){\line(-1,0){7}}\put(26,-2){\circle*{2}}
\put(19,-2){\circle*{2}}
\end{picture}}
\newcommand{\lcho}{\begin{picture}(20,10)
\put(0,3){\line(1,0){10}}
\put(10,3){\circle*{3}}\put(10,3){\line(1,3){4}}
\put(10,3){\line(1,-3){4}}\end{picture}}
\newcommand{\lchoo}{\begin{picture}(40,15)
\put(0,3){\line(1,0){10}} \put(15,3){\circle{10}}
\put(10,3){\circle*{3}}\put(20,3){\circle*{3}}
\put(20,3){\line(1,0){10}}
\put(30,3){\circle*{3}}\put(30,3){\line(1,3){4}}
\put(30,3){\line(1,-3){4}}\end{picture}}
\newcommand{\lchooo}{\begin{picture}(60,15)
\put(0,3){\line(1,0){10}} \put(15,3){\circle{10}}
\put(10,3){\circle*{3}}\put(20,3){\circle*{3}}
\put(20,3){\line(1,0){10}} \put(35,3){\circle{10}}
\put(30,3){\circle*{3}}\put(40,3){\circle*{3}}
\put(40,3){\line(1,0){10}}
\put(50,3){\circle*{3}}\put(50,3){\line(1,3){4}}
\put(50,3){\line(1,-3){4}}\end{picture}}
\newcommand{\lcholo}{\begin{picture}(60,15)
\put(0,3){\line(1,0){10}} \put(15,3){\circle{10}}
\put(10,3){\circle*{3}}\put(20,3){\circle*{3}}
\put(20,3){\line(1,0){20}}
\put(30,3){\circle*{3}}\put(30,3){\line(0,1){10}}
\put(45,3){\circle{10}}
\put(40,3){\circle*{3}}\put(50,3){\circle*{3}}
\put(50,3){\line(1,0){10}}
\end{picture}}
\newcommand{\lchllu}{\begin{picture}(60,15)
\put(0,3){\line(1,0){10}}
\put(25,16){\oval(10,10)[b]}
\put(10,3){\circle*{3}}\put(20,16){\circle*{3}}
\put(25,3){\circle{29}}
\put(30,16){\circle*{3}}\put(40,3){\circle*{3}}
\put(40,3){\line(1,0){10}}
\put(50,3){\circle*{3}}\put(50,3){\line(1,3){4}}
\put(50,3){\line(1,-3){4}}\end{picture}}
\newcommand{\lchlld}{\begin{picture}(60,15)
\put(0,3){\line(1,0){10}}
\put(25,-10){\oval(10,10)[t]}
\put(10,3){\circle*{3}}\put(20,-10){\circle*{3}}
\put(25,3){\circle{29}}
\put(30,-10){\circle*{3}}\put(40,3){\circle*{3}}
\put(40,3){\line(1,0){10}}
\put(50,3){\circle*{3}}\put(50,3){\line(1,3){4}}
\put(50,3){\line(1,-3){4}}\end{picture}}
\newcommand{\troodota}{\begin{picture}(35,26)
\put(0,3){\line(1,0){14}}
\put(14,3){\circle*{3}}\put(14,3){\line(1,1){20}}
\put(19,8){\circle*{3}}
\put(14,3){\line(1,-1){20}} \put(26,-8){\line(0,1){22}}
\put(26,-8){\circle*{3}}\put(26,14){\circle*{3}}
\end{picture}}
\newcommand{\troodotb}{\begin{picture}(35,15)
\put(0,3){\line(1,0){14}}
\put(14,3){\circle*{3}}\put(14,3){\line(1,1){20}}
\put(19,-2){\circle*{3}}
\put(14,3){\line(1,-1){20}} \put(26,-8){\line(0,1){22}}
\put(26,-8){\circle*{3}}\put(26,14){\circle*{3}}
\end{picture}}
\newcommand{\troodotc}{\begin{picture}(35,26)
\put(0,3){\line(1,0){14}}
\put(14,3){\circle*{3}}\put(14,3){\line(1,1){20}}
\put(26,3){\circle*{3}}
\put(14,3){\line(1,-1){20}} \put(26,-8){\line(0,1){22}}
\put(26,-8){\circle*{3}}\put(26,14){\circle*{3}}
\end{picture}}
\newcommand{\troodloop}{\begin{picture}(35,26)
\put(0,3){\line(1,0){14}}
\put(14,3){\circle*{3}}\put(14,3){\line(1,1){20}}
\put(26,3){\oval(8,8)[r]}\put(26,7){\circle*{3}}
\put(26,-1){\circle*{3}}
\put(14,3){\line(1,-1){20}} \put(26,-8){\line(0,1){22}}
\put(26,-8){\circle*{3}}\put(26,14){\circle*{3}}
\end{picture}}
\newcommand{\buble}{\begin{picture}(40,15)
\put(0,3){\line(1,0){10}} \put(18,3){\circle{16}}
\put(10,3){\circle*{3}}\put(27,3){\circle*{3}}
\put(28,3){\line(1,0){10}}
\end{picture}}
\newcommand{\bublel}{\begin{picture}(40,15)
\put(0,3){\line(1,0){10}} \put(18,3){\circle{16}}
\put(10,3){\circle*{3}}\put(27,3){\circle*{3}}
\put(28,3){\line(1,0){10}}
\put(18,-5){\line(0,1){16}}\put(18,-5){\circle*{3}}
\put(18,11){\circle*{3}}
\end{picture}}
\newcommand{\bubleldot}{\begin{picture}(40,15)
\put(0,3){\line(1,0){10}} \put(18,3){\circle{16}}
\put(10,3){\circle*{3}}\put(27,3){\circle*{3}}
\put(28,3){\line(1,0){10}}
\put(18,-5){\line(0,1){16}}\put(18,-5){\circle*{3}
}\put(18,11){\circle*{3}}
\put(12,9){\circle*{3}}
\end{picture}}
\newcommand{\bublelmidledot}{\begin{picture}(40,15)
\put(0,3){\line(1,0){10}} \put(18,3){\circle{16}}
\put(10,3){\circle*{3}}\put(27,3){\circle*{3}}
\put(28,3){\line(1,0){10}}
\put(18,-5){\line(0,1){16}}\put(18,-5){\circle*{3}}
\put(18,11){\circle*{3}}
\put(18,3){\circle*{3}}
\end{picture}}
\newcommand{\oloop}{\begin{picture}(12,12)
\put(0,-2){\line(1,1){10}}
\put(0,8){\line(1,-1){10}}
\put(5,3){\circle*{3}}
\end{picture}}
\newcommand{\sloop}{\begin{picture}(40,15)
\put(10,3){\line(-1,1){10}}\put(10,3){\line(-1,-1){10}}
\put(18,3){\circle{16}}
\put(10,3){\circle*{3}}\put(27,3){\circle*{3}}
\put(28,3){\line(1,1){10}}\put(28,3){\line(1,-1){10}}
\end{picture}}

\newcommand{\sloopovaled}{\begin{picture}(40,15)
\put(10,3){\line(-1,1){10}}\put(10,3){\line(-1,-1){10}}
\put(18,3){\circle{16}}
\put(10,3){\circle*{3}}\put(27,3){\circle*{3}}
\put(28,3){\line(1,1){10}}\put(28,3){\line(1,-1){10}}
\put(18,3){\oval(32,22)}
\end{picture}}
\newcommand{\sloopleftdot}{\begin{picture}(40,15)
\put(10,3){\line(-1,1){10}}\put(10,3){\line(-1,-1){10}}
\put(18,3){\circle{16}}\put(10,3){\circle*{3}}
\put(10,3){\circle{5}}\put(27,3){\circle*{3}}
\put(28,3){\line(1,1){10}}\put(28,3){\line(1,-1){10}}
\end{picture}}
\newcommand{\sloopleftdotovaled}{\begin{picture}(40,15)
\put(10,3){\line(-1,1){10}}\put(10,3){\line(-1,-1){10}}
\put(18,3){\circle{16}}\put(10,3){\circle*{3}}
\put(10,3){\circle{5}}\put(27,3){\circle*{3}}
\put(28,3){\line(1,1){10}}\put(28,3){\line(1,-1){10}}
\put(18,3){\oval(32,22)}
\end{picture}}
\newcommand{\slooprightdot}{\begin{picture}(40,15)
\put(10,3){\line(-1,1){10}}\put(10,3){\line(-1,-1){10}}
\put(18,3){\circle{16}}\put(10,3){\circle*{3}}
\put(27,3){\circle*{3}}\put(27,3){\circle{5}}
\put(28,3){\line(1,1){10}}\put(28,3){\line(1,-1){10}}
\end{picture}}
\newcommand{\slooprightdotovaled}{\begin{picture}(40,15)
\put(10,3){\line(-1,1){10}}\put(10,3){\line(-1,-1){10}}
\put(18,3){\circle{16}}\put(10,3){\circle*{3}}
\put(27,3){\circle*{3}}\put(27,3){\circle{5}}
\put(28,3){\line(1,1){10}}\put(28,3){\line(1,-1){10}}
\put(18,3){\oval(32,22)}
\end{picture}}
\newcommand{\tloop}{\begin{picture}(20,15)
\put(8,11){\line(-1,1){10}}\put(8,11){\line(1,1){10}}
\put(8,3){\circle{16}}
\put(8,11){\circle*{3}}\put(8,-5){\circle*{3}}
\put(8,-5){\line(-1,-1){10}}\put(8,-5){\line(1,-1){10}}
\end{picture}}
\newcommand{\uloop}{\begin{picture}(40,15)
\put(10,3){\line(-1,1){10}}
\put(29,3){\oval(38,30)[b,l]}
\put(18,3){\circle{16}}
\put(10,3){\circle*{3}}\put(27,3){\circle*{3}}
\put(28,3){\line(1,1){10}}
\put(8,3){\oval(38,30)[b,r]}
\end{picture}}
\newcommand{\siiloop}{\begin{picture}(50,15)
\put(10,3){\line(-1,1){10}}\put(10,3){\line(-1,-1){10}}
\put(10,3){\circle*{3}} \put(18,3){\circle{16}}
\put(27,3){\circle*{3}}
\put(35,3){\circle{16}}\put(43,3){\circle*{3}}
\put(43,3){\line(1,1){10}}\put(43,3){\line(1,-1){10}}
\end{picture}}
\newcommand{\siiloopovaled}{\begin{picture}(50,15)
\put(10,3){\line(-1,1){10}}\put(10,3){\line(-1,-1){10}}
\put(10,3){\circle*{3}} \put(18,3){\circle{16}}
\put(27,3){\circle*{3}}
\put(35,3){\circle{16}}\put(43,3){\circle*{3}}
\put(43,3){\line(1,1){10}}\put(43,3){\line(1,-1){10}}
\put(27,3){\oval(40,25)}
\end{picture}}
\newcommand{\siiloopboxed}{\begin{picture}(50,15)
\put(10,3){\line(-1,1){10}}\put(10,3){\line(-1,-1){10}}
\put(10,3){\circle*{3}} \put(18,3){\circle{16}}
\put(27,3){\circle*{3}}
\put(35,3){\circle{16}}\put(43,3){\circle*{3}}
\put(43,3){\line(1,1){10}}\put(43,3){\line(1,-1){10}}
\put(3,-7){\line(1,0){46}}\put(3,13){\line(1,0){46}}
\put(3,-7){\line(0,1){20}}\put(49,-7){\line(0,1){20}}
\end{picture}}
\newcommand{\siilooplboxed}{\begin{picture}(50,15)
\put(10,3){\line(-1,1){10}}\put(10,3){\line(-1,-1){10}}
\put(10,3){\circle*{3}} \put(18,3){\circle{16}}
\put(27,3){\circle*{3}}
\put(35,3){\circle{16}}\put(43,3){\circle*{3}}
\put(43,3){\line(1,1){10}}\put(43,3){\line(1,-1){10}}
\put(3,-7){\line(1,0){30}}\put(3,13){\line(1,0){30}}
\put(3,-7){\line(0,1){20}}\put(33,-7){\line(0,1){20}}
\end{picture}}
\newcommand{\siiloopalboxed}{\begin{picture}(50,15)
\put(10,3){\line(-1,1){10}}\put(10,3){\line(-1,-1){10}}
\put(10,3){\circle*{3}} \put(18,3){\circle{16}}
\put(27,3){\circle*{3}}
\put(35,3){\circle{16}}\put(43,3){\circle*{3}}
\put(43,3){\line(1,1){10}}\put(43,3){\line(1,-1){10}}
\put(5,-7){\line(1,0){28}}\put(5,13){\line(1,0){28}}
\put(5,-7){\line(0,1){20}}\put(33,-7){\line(0,1){20}}
\put(2,-10){\line(1,0){47}}\put(2,16){\line(1,0){47}}
\put(2,-10){\line(0,1){26}}\put(49,-10){\line(0,1){26}}
\end{picture}}
\newcommand{\siilooprboxed}{\begin{picture}(50,15)
\put(10,3){\line(-1,1){10}}\put(10,3){\line(-1,-1){10}}
\put(10,3){\circle*{3}} \put(18,3){\circle{16}}
\put(27,3){\circle*{3}}
\put(35,3){\circle{16}}\put(43,3){\circle*{3}}
\put(43,3){\line(1,1){10}}\put(43,3){\line(1,-1){10}}
\put(21,-7){\line(1,0){28}}\put(21,13){\line(1,0){28}}
\put(21,-7){\line(0,1){20}}\put(49,-7){\line(0,1){20}}
\end{picture}}
\newcommand{\siilooparboxed}{\begin{picture}(50,15)
\put(10,3){\line(-1,1){10}}\put(10,3){\line(-1,-1){10}}
\put(10,3){\circle*{3}} \put(18,3){\circle{16}}
\put(27,3){\circle*{3}}
\put(35,3){\circle{16}}\put(43,3){\circle*{3}}
\put(43,3){\line(1,1){10}}\put(43,3){\line(1,-1){10}}
\put(21,-7){\line(1,0){26}}\put(21,13){\line(1,0){26}}
\put(21,-7){\line(0,1){20}}\put(47,-7){\line(0,1){20}}
\put(2,-10){\line(1,0){49}}\put(2,16){\line(1,0){49}}
\put(2,-10){\line(0,1){26}}\put(51,-10){\line(0,1){26}}
\end{picture}}
\newcommand{\siirloop}{\begin{picture}(33,15)
\put(10,3){\line(-1,1){10}}\put(10,3){\line(-1,-1){10}}
\put(18,3){\circle{16}} \put(10,3){\circle*{3}}
\put(20,-4){\line(0,1){14}}
\put(20,-4){\circle*{3}}\put(20,10){\circle*{3}}
\put(30,-4){\oval(20,10)[b,l]}\put(30,10){\oval(20,10)[t,l]}
\end{picture}}
\newcommand{\siirloopboxed}{\begin{picture}(33,15)
\put(10,3){\line(-1,1){10}}\put(10,3){\line(-1,-1){10}}
\put(18,3){\circle{16}}
\put(10,3){\circle*{3}}
\put(20,-4){\line(0,1){14}}
\put(20,-4){\circle*{3}}\put(20,10){\circle*{3}}
\put(30,-4){\oval(20,10)[b,l]}\put(30,10){\oval(20,10)[t,l]}
\put(3,-10){\line(1,0){24}}\put(3,16){\line(1,0){24}}
\put(3,-10){\line(0,1){26}}\put(27,-10){\line(0,1){26}}
\end{picture}}
\newcommand{\siirlooprboxed}{\begin{picture}(33,15)
\put(10,3){\line(-1,1){10}}\put(10,3){\line(-1,-1){10}}
\put(18,3){\circle{16}}
\put(10,3){\circle*{3}}
\put(20,-4){\line(0,1){14}}
\put(20,-4){\circle*{3}}\put(20,10){\circle*{3}}
\put(30,-4){\oval(20,10)[b,l]}\put(30,10){\oval(20,10)[t,l]}
\put(16,-8){\line(1,0){11}}\put(16,13){\line(1,0){11}}
\put(16,-8){\line(0,1){21}}\put(27,-8){\line(0,1){21}}
\end{picture}}
\newcommand{\siirlooparboxed}{\begin{picture}(33,15)
\put(10,3){\line(-1,1){10}}\put(10,3){\line(-1,-1){10}}
\put(18,3){\circle{16}}
\put(10,3){\circle*{3}}
\put(20,-4){\line(0,1){14}}
\put(20,-4){\circle*{3}}\put(20,10){\circle*{3}}
\put(30,-4){\oval(20,10)[b,l]}\put(30,10){\oval(20,10)[t,l]}
\put(16,-8){\line(1,0){11}}\put(16,13){\line(1,0){11}}
\put(16,-8){\line(0,1){21}}\put(27,-8){\line(0,1){21}}
\put(3,-10){\line(1,0){27}}\put(3,16){\line(1,0){27}}
\put(3,-10){\line(0,1){26}}\put(30,-10){\line(0,1){26}}
\end{picture}}
\newcommand{\siilloop}{\begin{picture}(40,15)
\put(26,3){\line(1,1){10}}\put(26,3){\line(1,-1){10}}
\put(18,3){\circle{16}}
\put(26,3){\circle*{3}}
\put(16,-4){\line(0,1){14}}
\put(16,-4){\circle*{3}}\put(16,10){\circle*{3}}
\put(6,-4){\oval(20,10)[b,r]}\put(6,10){\oval(20,10)[t,r]}
\end{picture}}
\newcommand{\contracted}{\begin{picture}(10,10)
\put(5,3){\circle{5}}\put(5,3){\circle*{3}}
\end{picture}}
\newcommand{\six}{\begin{picture}(10,12)
\put(5,3){\circle*{3}}
\put(5,-3){\line(0,1){12}}
\put(0,8){\line(1,-1){10}}
\put(0,-2){\line(1,1){10}}
\end{picture}}
\newcommand{\eight}{\begin{picture}(10,12)
\put(5,3){\circle*{3}}\put(-1,3){\line(1,0){12}}
\put(5,-3){\line(0,1){12}}
\put(0,8){\line(1,-1){10}}
\put(0,-2){\line(1,1){10}}
\end{picture}}
\newcommand{\sixloop}{\begin{picture}(25,15)
\put(-2,0){\line(1,0){29}}\put(5,0){\circle*{3}}
\put(0,-5){\line(1,1){17}}\put(20,0){\circle*{3}}
\put(25,-5){\line(-1,1){17}}\put(13,8){\circle*{3}}
\end{picture}}
\newcommand{\sixloopbuble}{\begin{picture}(25,15)
\put(-2,0){\line(1,0){29}}\put(5,0){\circle*{3}}
\put(0,-5){\line(1,1){13}}\put(20,0){\circle*{3}}
\put(25,-5){\line(-1,1){13}}\put(13,8){\circle*{3}}
\put(13,12){\circle{8}}\put(13,16){\circle*{3}}
\put(13,16){\line(6,1){10}}\put(13,16){\line(-6,1){10}}
\end{picture}}
\newcommand{\sixbuble}{\begin{picture}(40,15)
\put(10,3){\line(-2,1){10}}\put(10,3){\line(-2,-1){10}}
\put(10,3){\line(-1,1){10}}\put(10,3){\line(-1,-1){10}}
\put(18,3){\circle{16}}
\put(10,3){\circle*{3}}\put(27,3){\circle*{3}}
\put(28,3){\line(1,1){10}}\put(28,3){\line(1,-1){10}}
\end{picture}}
\newcommand{\eightbuble}{\begin{picture}(40,15)
\put(10,3){\line(-2,1){10}}\put(10,3){\line(-2,-1){10}}
\put(10,3){\line(-1,1){10}}\put(10,3){\line(-1,-1){10}}
\put(18,3){\circle{16}}
\put(10,3){\circle*{3}}\put(27,3){\circle*{3}}
\put(28,3){\line(1,1){10}}\put(28,3){\line(1,-1){10}}
\put(28,3){\line(2,1){10}}\put(28,3){\line(2,-1){10}}
\end{picture}}
\newcommand{\sixIbuble}{\begin{picture}(40,24)
\put(10,8){\line(-1,-1){10}}\put(10,8){\line(-1,1){10}}
\put(18,8){\circle{16}}
\put(10,8){\circle*{3}}\put(27,8){\circle*{3}}
\put(28,8){\line(1,1){10}}\put(28,8){\line(1,-1){10}}
\put(18,-7){\line(0,1){30}}
\put(18,0){\circle*{3}}\put(18,16){\circle*{3}}
\end{picture}}
\newcommand{\cross}{\begin{picture}(15,10)
\put(0,3){\line(1,0){14}}
\put(3,0){$\times$}
\end{picture}}
\newcommand{\onepi}{\begin{picture}(32,24)
\put(-1,3){\line(1,0){31}}
\put(14,3){\circle{15}}
\put(7,3){\circle*{3}}\put(22,3){\circle*{3}}
\end{picture}}
\newcommand{\twopi}{\begin{picture}(30,20)
\put(21,3){\line(1,0){7}}
\put(14,3){\circle{15}}\put(35,3){\circle{15}}
\put(11,1){$\gamma_1$}\put(32,1){$\gamma_2$}
\end{picture}}
\newcommand{\oneloopmass}{\begin{picture}(20,24)
\put(0,3){\line(1,0){18}}\put(9,7){\circle{8}}
\put(9,3){\circle*{3}}
\end{picture}}
\newcommand{\twoloopmass}{\begin{picture}(20,24)
\put(0,3){\line(1,0){18}}\put(9,7){\circle{8}}
\put(9,3){\circle*{3}}\put(9,15){\circle{8}}
\put(9,11){\circle*{3}}
\end{picture}}
\newcommand{\sloopm}{\begin{picture}(40,15)
\put(10,3){\line(-1,1){10}}\put(10,3){\line(-1,-1){10}}
\put(18,3){\circle{16}}\put(18,15){\circle{8}}
\put(18,11){\circle*{3}}\put(10,3){\circle*{3}}
\put(27,3){\circle*{3}}\put(28,3){\line(1,1){10}}
\put(28,3){\line(1,-1){10}}
\end{picture}}

\hoffset= - 1.0in         

\begin{document}

\hfill ITEP/TH-08/07

\bigskip

\centerline{\Large{
Non-Linear Algebra and Bogolubov's Recursion
}}

\bigskip

\centerline{\it A.Morozov}

\bigskip

\centerline{ITEP, Moscow, Russia}

\bigskip

\centerline{\it M.Serbyn\footnote{Also at Moscow Physical
technical Institute, Landau Institute of Theoretical Physics and
Kiev Institute of Theoretical Physics \\
e-mail: maksym.serbyn@itep.ru} }

\bigskip

\centerline{Museo Storico della Fisica e Centro Studi e Ricerche
"Enrico Fermi", Rome, Italy}
\centerline{ITEP, Moscow, Russia}

\bigskip

\centerline{ABSTRACT}

\bigskip

Numerous examples are given of application of Bogolubov's
forest formula to iterative solutions of various non-linear
equations:
one and the same formula describes everything,
from ordinary quadratic equation to renormalization
in quantum field theory.

\bigskip

\bigskip

\tableofcontents

\section{Introduction}

Bogolubov's forest formula \cite{Bogo}-\cite{Vass}
is one of the gemstones
of Quantum Field Theory (QFT), used to handle iterative
renormalization procedure.
In \cite{CK} and especially in \cite{GMS} it was shown that
forest formula is actually a pure algebraic construction,
devised to solve iteratively a broad set of non-linear equation
and, remarkably, one and the same formula describes everything:
from solution to an ordinary quadratic equation to
Lamb shift and renormalization of the Standard Model.
The general presentation of ref.\cite{GMS} can be a little
too abstract and does not contain examples.
Some examples can be found in \cite{CK2}-\cite{Vol}, but
they are all at the level of quantum field theory and therefore
obscure the simple algebraic and tensorial structure.
A pedagogical review \cite{Dela} deals with adequately simple
examples, but it does not pay enough attention to algebra and
forest formula.
It is a purpose of this paper to provide the needed illustrations
to the formalism of \cite{GMS}, and this is a natural addendum to
generic presentation of non-linear algebra in ref.\cite{nolal}.

\subsection{The problem}

Forest formula provides iterative solution to the following problem.

$\bullet$
Let $F(T)$ be a {\it given} function of some variables $T$,
called times or coupling constants.
In QFT applications $F(T)$ is usually an effective action
(logarithm of partition function) of a physical theory,
considered as a function of {\it bare} couplings, but this
interpretation is inessential for the forest formula {\it per se}.

$\bullet$
Let ${\cal P}_-$ be a linear projector onto the linear sub-space of
{\it unwanted} functions.
In renormalization-theory context the function $F(T)$ depends also
on a cut-off $\Lambda$, and {\it unwanted} are functions, which are
singular in the limit $\Lambda \rightarrow \infty$.
However, again, concrete interpretation of projector ${\cal P}_-$
is inessential for the forest formula.

$\bullet$
Given $F(T)$ and ${\cal P}_-$ we are ready to formulate the
{\bf problem}.
Can one make a change of {\it arguments} $T$ in $F(T)$, without
changing the {\it shape} of this function, to put it into the
sub-space of {\it wanted} functions?
In other words, the problem is to find a change of variables
$T \rightarrow \tilde T = T + Q(T)$, such that
\be
{\cal P}_-\left\{ F\Big(T + Q(T)\Big)\right\} = 0
\label{prob1}
\ee
To make solution of this problem unambiguous, one can impose
additional constraint that the {\it counter-term} $Q(T)$
is pure singular, i.e. lies entirely in the space of {\it unwanted}
functions:
\be
{\cal P}_+\Big\{Q(T)\Big\} = 0,
\label{prob2}
\ee
where ${\cal P}_+ = I - {\cal P}_-$ is projector onto the space
of {\it wanted} functions\footnote{In fact,
this conditions is somewhat more complicated:
${\cal P}_+\Big\{\hat F(\Gamma/\Gamma)\hat Q(\Gamma)\Big\} = 0$,
see sections \ref{241} and \ref{242} \label{footnote}}.

\subsection{Forest formula: solution to the problem in terms of
Feynman diagrams}

Forest formula provides an explicit solution to the problem
(\ref{prob1}) \& (\ref{prob2}), and the solution is in terms of
graph theory!
The crucial observation \cite{GMS} is that any formal series
in non-negative powers of couplings $T$ can be represented as
a sum over Feynman diagrams, with $T$ standing at the vertices,
and this is the most adequate language for tensor-algebra
studies \cite{nolal}, see examples below.
This means that both the $F(T)$ and the counter-term
$Q(T)$ can be expanded into sums over graphs:
\be
F(T) = \sum_\Gamma \hat F(\Gamma) Z(\Gamma|T), \ \ \ \
Q(T) = \sum_\Gamma \hat Q(\Gamma) Z(\Gamma|T)
\ee
with basic functions $Z(\Gamma|T)$, capturing the topology
of the diagram $\Gamma$: $Z(\Gamma|T)$ is a product of
couplings $T$ at the vertices with indices (if any)
contracted along the links.
We assume that projectors ${\cal P}_\pm$ act on the
coefficients $\hat F(\Gamma)$ and $\hat Q(\Gamma)$ rather
than on $Z(\Gamma|T)$, and this is the case in most applications.

Now iterative solution to the problem
(\ref{prob1}) \& (\ref{prob2})
is given by the {\it forest formula}:
\be
\hat F(\Gamma/\Gamma) \hat Q(\Gamma) = -{\cal P}_-\left\{
\hat F(\Gamma) +
\sum_{\big\{\gamma_1 \cup \ldots \cup \gamma_k\big\}}
\hat F\Big(\Gamma/\gamma_1\ldots \gamma_k\Big)
\hat Q(\gamma_1)\ldots\hat Q(\gamma_k)\right\}
\label{fof}
\ee
where $\big\{\gamma_1 \cup \ldots \cup \gamma_k\big\}$
are all possible box-subgraphs in graph
$\Gamma$, $\gamma_i\in{\cal B}\Gamma$,
i.e. parts of
$\Gamma$ lying in a collection of non-intersecting "boxes",
and $\Gamma/\gamma_1\ldots \gamma_k$ is obtained by contracting
all boxes to points.
The factor $\Gamma/\Gamma$ is just a single vertex with the same
external lines as original $\Gamma$, and $\hat Q(\Gamma)\neq 0$
and is given by eq.(\ref{fof}) only if
$\hat F(\Gamma/\Gamma)\neq 0$,
i.e. depending on the shape of the function $F(T)$.
The meaning of this criterium is simple: it is the coupling at
this vertex $\hat F(\Gamma/\Gamma)$ that is shifted by $\hat Q(\Gamma)$.
We ascribe a special name "vertex criterium" to this rule,
because of its significance for our further considerations.

We refer to \cite{CK,GMS} for discussion of algebraic theory
behind eq.(\ref{fof}), its relation to the group of
diffeomorphisms of the moduli space of theories and to its dual
Connes-Kreimer Hopf algebra on graphs. Instead in what follows we
consider various examples and applications of eq.(\ref{fof}).

\subsection{The second-level forest formula \label{intro}}

In addition to the first-level forest formula (\ref{fof}),
there is a next-level Bogolubov forest formula
\cite{Bogo}-\cite{Peskin},
involving sums over embedded box-subgraphs,
see eq.(9.7) of \cite{GMS}.
It gives directly the answer for the renormalized effective action
$F(\tT)=F(T+Q(T))=\sum_\Gamma \hat F_R(\Gamma) Z(\Gamma|T)$:
\be
\hat F_R(\Gamma) = \sum_{{\cal F}_\Gamma}
\prod_{{\cal T}\in{\cal F}_\Gamma}
\left( \prod_{\stackrel{vertices}{of\ {\cal T}}}^{\longrightarrow}
\frac{1}{\hat F(\hat\gamma_k/\hat\gamma_k)}(-{\cal P}_-)
\hat F\left(\hat\gamma_k/(\hat\gamma_{k+1}^1\cdot\ldots\cdot
\hat\gamma_{k+1}^{s(k)})\right)
\right)
\label{fof2}
\ee
The sum is taken over all possible forests ${\cal F}_\Gamma$,
which are unions of trees for every connected component of $\Gamma$.
For some connected component of  $\Gamma$, $\Gamma_1$,
the tree ${\cal T}$ is build in the following way:

\textbf{i}) Choose concrete sequence of embedded box-subgraphs
$\{\gamma_n\}$:
\be
 \gamma_0 = \Gamma_1,\ \gamma_1 \in
{\cal B}\Gamma,\ \gamma_2 \in {\cal B}\gamma_1 \subset
{\cal B}\Gamma, \ldots, \gamma_n \in {\cal B}\gamma_{n-1} \subset
\ldots \subset {\cal B}\gamma_1 \subset {\cal B}\Gamma_1
\label{seq_gamma}\ee
where graphs $\gamma_k$ are not obligatory connected
and can consist of $s(k)$ connected components
$\gamma_k^{1},\ldots,\gamma_k^{s(k)}$.
But for every $\gamma_k^{i}$ should exist such $\gamma_{k-1}^{j(i)}$,
that $\gamma_k^{i} \in {\cal B}\gamma_{k-1}^{j(i)}$
(condition (\ref{seq_gamma}) is understood exactly in this way).

\textbf{ii}) Making use of this sequence of embedded
box-subgraphs, or, equivalently, of the collection of non-intersecting
boxes, build a decorated rooted tree ${\cal T}$.
With lower site of each box one associates a vertex of the tree.
Two vertices are connected by a link,
if one of the corresponding boxes lies
immediately inside another
(i.e. there are no boxes in between the two).
The root link ends at a vertex,
associated with $\gamma_0 = \Gamma$.
In terms of graphs, now a vertex of the tree is associated
with every $\gamma_k^{i}$,
and there is exactly one link which goes downwards
(towards the root, i.e. associated with the neighboring bigger box)
and connects $\gamma_k^i$ to $\gamma_{k-1}^{j(i)}$,
and unrestricted number of links, going upwards and connecting
$\gamma_k$ to some collection
$\gamma_{k+1}^1,\ldots, \gamma_{k+1}^{s(k,i)} \subset \gamma_{k+1}$.

The tree product
$$\prod_{\stackrel{vertices}{of\ {\cal T}}}^{\longrightarrow}$$
associates with every vertex
$\gamma_k^i$ an operator
$\left(\hat F\big([\gamma_k^i/\gamma_k^i]\big)\right)^{-1}
(-{\cal P}_-) \hat F\left(\gamma_k^i/(\gamma_{k+1}^1\cdot\ldots\cdot
\gamma_{k+1}^{j(i)})\right)$,
while projector ${\cal P}_-$
acts upwards along the branches of the tree.
The root vertex $\gamma_0$ (i.e. a connected component of $\Gamma$)
contributes just
$ \hat F\Big(\gamma_0/(\gamma_{1}^1\cdot\ldots\cdot
\gamma_{1}^{s(0)})\Big)$.
Arrow over the product sign means that the product
is ordered along the branches.

\bigskip

\bigskip

The problem (\ref{prob1}) \& (\ref{prob2}) implies that
function $F(T)$ and projector ${\cal P}_-$ are already given.
However, the diagram technique still needs to be introduced.
Moreover, for given $F(T)$ and ${\cal P}_-$ this can be done
in different ways.
It deserves mentioning also that iterative nature of forest
formula does not mean that there is a {\it small} expansion
parameter: moreover, in renormalization-theory context the
relevant expansion parameter is rather large than small.

\part{Algebraic examples}

\section{Quadratic equation.
A valence-2 vertex and tree diagrams}

Our starting example is
\be
\left\{ \begin{array}{c}
F(T) = T + \epsilon T^2, \\
{\cal P}_-\Big\{f(\epsilon)\Big\} = f(\epsilon) - f(0),
\end{array} \right.
\label{trivproj} \ee i.e. "unwanted" are functions, which depend
on $\epsilon$ and the goal is to eliminate $\epsilon$-dependence
by $\epsilon$-dependent shift of variables $T \rightarrow \tilde T
+ Q_\epsilon(T)$, so that $F(\tilde T) = \tilde T + \epsilon
\tilde T^2$ no longer depends on $\epsilon$. Imposing the
constraint (\ref{prob2}), we reduce the problem to \be \tilde T +
\epsilon \tilde T^2 = T \ee and therefore \be \tilde T =
\frac{\sqrt{1 + 4\epsilon T}-1}{2\epsilon} = T+ \sum_{k=1}
\frac{(-)^{k}2^{k}(2k-1)!!}{(k+1)!}\epsilon^{k}T^{k+1}
 = T - \epsilon T^2 + 2\epsilon^2 T^3 - 5\epsilon^3 T^4 +
14\epsilon^4 T^5 - \ldots  \label{soquad} \ee

Now we are going to reproduce the same series from the forest formula
(\ref{fof}).
To do this, we need first to choose a particular diagram technique.
Our first choice will be a single valence-$2$ vertex $T = T_{ij}$,
where indices $i,j$ take a single value $i,j=1$
and will be omitted at the beginning.
The only possible connected Feynman diagrams are chains and circles,
both equal to $T^n$, where $n$ is the number of vertices.
Only two diagrams, with $n=1$ and $n=2$ contribute to $F(T)$.

\subsection{Chain diagrams}

Assume first, that these are two chain diagrams:\footnote{In
what follows we omit hats over $\hat Q(\Gamma)$ and
$\hat F(\Gamma)$, assuming that they will not be confused
with the corresponding functions $Q(T)$ and $F(T)$ of
$T$-variables.}
\be
\hat F(\go)=1, \ \ \ \  \hat F(\goo)=\epsilon
\ee
Accordingly $F(\Gamma/\Gamma)\neq 0$
only for chain diagrams $\Gamma$.
By {\it vertex criterium} this implies that the only
non-vanishing are contributions to $Q$ of chain-diagrams.
For them the forest formula (\ref{fof}) gives:
$$
\begin{array}{l}
Q[1] \equiv Q(\go) = -{\cal
P}_-\Big\{F(\go)\Big\} = 0,  \\
Q[2] \equiv Q(\goo) = -{\cal
P}_-\Big\{F(\goo)\Big\} = -\epsilon,  \\
Q[3] \equiv Q(\gooo) =
-{\cal P}_-\Big\{F(\gooo) + 2F(\goo)Q(\goo)\Big\} = \\
\ \ \ \ \ \ \ = -2{\cal P}_-\Big\{F(\goo)Q(\goo)\Big\} =
+2\epsilon^2
\end{array}
$$
In the last formula $F(\gooo) = 0$ because of our choice of
$F(T)$, and coefficient $2$ in from of the next term appears
because there are two different possibilities to choose a
box-subgraph \goo \ inside \ \gooo:
\ \ \iiiboxleft and \iiiboxright.

Similarly, three box-subgraphs

\begin{picture}(42,10) \put(0,3){\line(1,0){40}}
\put(5,2){\circle*{3}}\put(15,2){\circle*{3}}
\put(25,2){\circle*{3}}\put(35,2){\circle*{3}}
\put(3,-2){\line(0,1){8}} \put(3,6){\line(1,0){28}}
\put(31,-2){\line(0,1){8}}\put(3,-2){\line(1,0){28}}
\end{picture},
\begin{picture}(42,10) \put(0,3){\line(1,0){40}}
\put(5,2){\circle*{3}}\put(15,2){\circle*{3}}
\put(25,2){\circle*{3}}\put(35,2){\circle*{3}}
\put(3,-2){\line(0,1){8}} \put(3,6){\line(1,0){16}}
\put(19,-2){\line(0,1){8}}\put(3,-2){\line(1,0){16}}
\put(21,-2){\line(0,1){8}} \put(21,6){\line(1,0){18}}
\put(39,-2){\line(0,1){8}}\put(21,-2){\line(1,0){18}}
\end{picture}
and
\begin{picture}(42,10) \put(0,3){\line(1,0){40}}
\put(5,2){\circle*{3}}\put(15,2){\circle*{3}}
\put(25,2){\circle*{3}}\put(35,2){\circle*{3}}
\put(10,-2){\line(0,1){8}} \put(10,6){\line(1,0){28}}
\put(38,-2){\line(0,1){8}}\put(10,-2){\line(1,0){28}}
\end{picture}
contribute to the next $Q$:
$$
Q[4] \equiv
Q(\goooo) = -{\cal P}_-\left\{F(\goo) \left(2Q(\gooo) +
\big[Q(\goo)\big]^2\right)\right\} = -5\epsilon^3
$$
Note that box-subgraphs
\begin{picture}(42,10) \put(0,3){\line(1,0){40}}
\put(5,2){\circle*{3}}\put(15,2){\circle*{3}}
\put(25,2){\circle*{3}}\put(35,2){\circle*{3}}
\put(10,-2){\line(0,1){8}} \put(10,6){\line(1,0){18}}
\put(28,-2){\line(0,1){8}}\put(10,-2){\line(1,0){18}}
\end{picture}
and
\begin{picture}(42,10) \put(0,3){\line(1,0){40}}
\put(5,2){\circle*{3}}\put(15,2){\circle*{3}}
\put(25,2){\circle*{3}}\put(35,2){\circle*{3}}
\put(3,-2){\line(0,1){8}} \put(3,6){\line(1,0){18}}
\put(21,-2){\line(0,1){8}}\put(3,-2){\line(1,0){18}}
\end{picture}
 do not contribute, because
contraction of boxes produces {\it three}-vertex diagrams and such
contributions would be multiplied by $F(\gooo)$, which vanishes
for our choice of $F(T)$. For continuation of the calculation let
us denote the $N$-vertex chain diagram through $[N]$. Then
$$
Q[5] = -{\cal P}_-\left\{F[2]\Big(2Q[4] + 2Q[2]Q[3]\Big)\right\} =
14\epsilon^4,
$$
and, in general, eq.(\ref{fof}) implies
$$
Q[N] = -{\cal P}_-\left\{F[2]
\left(\delta_{N,2} + 2Q[N-1]+
\sum_{k=2}^{N-2}Q[k]Q[N-k]\right)\right\},
\ \ N\geq 3
$$
For generating function $Q(t) = \sum_{N=2}t^NQ[N]$ this gives:
\be
Q = -F[2] \Big(t^2+ 2tQ + Q^2\Big) \label{summQ}
\ee
with $F[2] =
\epsilon$, i.e. our familiar (\ref{soquad}):
\be  Q(t)+t + \epsilon
(Q+t)^2=t, \ \ Q(t) + t = \frac{\sqrt{1+4\epsilon t}-1}{2\epsilon}
\ee

\subsection{Introducing coefficients}

In above example the basic functions $Z(\Gamma|T)$ entered
{\it original} function $F(T)$ with coefficients $1$ and $\epsilon$.
If instead we write
\be
F(T) = a\ \begin{picture}(20,10)\put(0,3){\line(1,0){20}}
\put(10,3){\circle*{3}}\end{picture} +
\epsilon b\ \begin{picture}(20,10)\put(0,3){\line(1,0){20}}
\put(5,3){\circle*{3}}\put(15,3){\circle*{3}}\end{picture}
= aT + \epsilon bT^2,
\ \ \ {\rm i.e.} \ \ \
\hat F(\go)=a, \ \ \ \  \hat F(\goo)=\epsilon b,
\ee
then the coefficient $a$ will appear at the l.h.s. of eq.(\ref{fof}):
$$
\begin{array}{l}
a Q(\go) = -{\cal P}_-\Big\{F(\go)\Big\} = 0,  \\
a Q(\goo) = -{\cal P}_-\Big\{F(\goo)\Big\} =
-\epsilon b,  \\
a Q(\gooo) = -2{\cal P}_-\Big\{F(\goo)Q(\goo)\Big\} =
\frac{2\epsilon^2 b^2}{a}
\end{array}
$$
so that
$$
\tilde T = T + Q(\go)T + Q(\goo)T^2 + Q(\gooo) T^3 + \ldots = T -
\frac{\epsilon b T^2}{a} + \frac{2\epsilon^2 b^2T^3}{a^2} - \ldots
$$
and
\be
F(\tilde T) = a\tilde T + \epsilon b \tilde T^2 =
a\left(T - \frac{\epsilon b T^2}{a} +
\frac{2\epsilon^2 b^2T^3}{a^2} + \ldots\right) + \epsilon b
\left(T - \frac{\epsilon b T^2}{a} +
\frac{2\epsilon^2 b^2T^3}{a^2} + \ldots\right)^2 = aT
\label{1234}
\ee

\subsection{Application of the second-level forest formula }

The second-level forest formula (\ref{fof2}) produces
the renormalized function $F(\tilde T)$ directly.
we already know what it is from eq.(\ref{1234}) and
this answer is very simple:
$$F(\tT)=F_R(T)=aT$$
This means that in this case the only non-vanishing
output of eq.(\ref{fof2}) should be $F_R(\go)=a$,
while for all other graphs $F_R(\Gamma)=0$.
The first statement is immediately reproduced by (\ref{fof2}):
there is only one terms at the r.h.s. of (\ref{fof2}).
For more complicated $\Gamma$ exact cancelation should occur
between different contributions to (\ref{fof2}).

Let us demonstrate this for \ \goo,
where one can build the following two trees:
$$\goo \ \ \mbox{and} \ \ \iibox$$
We do not consider trees which have box-subgraphs with one vertex
inside, like \ibox, because $\Pm F(\go)=0$, also we
do not draw external box which corresponds to the net graph.
Therefore
\be
F_R(\goo)=F(\goo)+F(\go)\frac{1}{F(\go)}(-\Pm)F(\goo)=
F(\goo)-\Pm F(\goo)=0
\label{fof2_goo}
\ee
Next is \gooo with embedded box-subgraphs
(only sub-trees that give non-vanishing contributions to
(\ref{fof2}) are shown):
\iiiboxleft, \iiiboxleftnet\  and\  \iiiboxright, \iiiboxrightnet.
It is quite obvious that the first two and the second two trees
give the same contribution, thus we take into account
only the first two graphs and
multiply the result by a factor of $2$:
\be
F_R(\gooo)= \nn \\
= 2\left\{F(\goo)\frac{1}{F(\go)}(-\Pm)F(\goo)+
F(\go)\frac{1}{F(\go)}(-\Pm)
\left[F(\goo)\frac{1}{F(\go)}(-\Pm)F(\goo)\right]\right\}
\label{fof2_gooo}
\ee
Thus $F_R(\gooo)$ vanishes,
\be
F_R(\gooo)=-\frac{2}{F(\go)}\left\{F(\goo)\Pm F(\goo)-
\Pm F(\goo)\Pm F(\goo)\right\}=0, \nn
\ee
because $F(\goo)$ is ${\cal P}$-negative.

Now it is easy to understand the way in which cancelations occur:
all trees for some particular graph can be divided into pairs:
the only difference between trees in one pair is that one of
them has "external" box, i.e. in terms of (\ref{fof2}),
$\gamma_0=\gamma_1=\Gamma_1$.
Due to the presence of additional projector $(-\Pm)$,
the two graphs in each pair exactly cancel each
other.

\bigskip

In what follows, we do not give examples
of application of second level forest formula in cases
where it is trivial,
because there the same argument can be used to proof
cancelation all except some peculiar graphs.

\subsection{Generic projector}

As the next generalization we do not fix concrete
$\epsilon$-dependence of the coefficients $A(\epsilon)$ and
$B(\epsilon)$ in
\be
F(T) = A(\epsilon)T + B(\epsilon)T^2 =
A(\epsilon)\ \go + B(\epsilon)\ \goo,
\ \ \ \ {\rm i.e.}\ \ \
\hat F(\go)=A(\epsilon), \ \ \  \hat F(\goo)=B(\epsilon)
\ee
(e.g., they can be generic Laurent series),
and concrete choice of projector ${\cal P}_-$
(e.g., it can pick up all terms with $\epsilon^k$ and
$k>k_0$ or, instead, all terms with $\ \epsilon^k\sinh\epsilon\ $
or anything else).
Eq.(\ref{fof}) works in all situations! Indeed, the
counter-terms are:
$$
A(\epsilon)Q\left(\go\right) = -{\cal P}_-\left\{
F\left(\go\right) \right\} = -A_-,
$$
i.e.
$$
Q\left(\go\right) = -\frac{A_-}{A}
$$
$$
A(\epsilon)Q\left(\goo\right) = -{\cal P}_-\left\{
F\left(\goo\right) \left(1 + 2Q\left(\go\right) +
\left[Q\left(\go\right) \right]^2\right)\right\} =
$$
$$
=-\left[B\left(1-\frac{A_-}{A}\right)^2\right]_- =
-\left(\frac{BA_+^2}{A^2}\right)_-
$$
$$
Q[3]=Q\left(\gooo\right)= -\frac{1}{A}{\cal
P}_-\left\{F[2]\left(2Q[2]+2Q[2]Q[1]\right)\right\}=
$$
$$
=\frac{2}{A} \left(B\frac{1}{A}\left(\frac{BA_+^2}{A^2}
\right)_-\left\{1-\frac{A_-}{A}\right\}\right)_-=\frac{2}{A}
\left(\frac{BA_+}{A^2}\left(\frac{BA_+^2}{A^2}\right)_-\right)_-
$$
$$
Q[4]=-\frac{1}{A}{\cal
P}_-\left(F[2]Q^2[2]+2F[2]Q[3]+2F[2]Q[3]Q[1]\right)=
$$
$$
=-\frac{1}{A}\left(\frac{B}{A^2}\left(\frac{BA_+^2}{A^2}\right)^2_-
+\frac{4BA_+}{A^2}\left(\frac{BA_+}{A^2}
\left(\frac{BA_+^2}{A^2}\right)_-\right)_-\right)_-
$$
$$ \ldots $$
Therefore
$$
\tilde T = A_+\frac{T}{A} -
\left(\frac{BA_+^2}{A^2}\right)_-\frac{T^2}{A} +
2\left(\frac{BA_+}{A^2}
\left(\frac{BA_+^2}{A^2}\right)_-\right)_-\frac{T^3}{A}
-
$$
$$
- \left\{4\left(\frac{BA_+}{A^2}
\left(\frac{BA_+}{A^2}\left(\frac{BA^2_+}{A^2}\right)_-\right)_
-\right)_-
+
\left(\frac{B}{A^2}\left(\frac{BA^2_+}{A^2}\right)_-^2\right)_-
\right\}
\frac{T^4}{A} + \ldots
$$
and
$$
F(\tilde T) = A\tilde T + B\tilde T^2 = A_+T +
\left(\frac{BA^2_+}{A^2}-\left(\frac{BA^2_+}{A^2}\right)_-\right)
T^2- 2\left[\frac{BA_+}{A^2}\left(\frac{BA^2_+}{A^2}\right)_- -
\left(\frac{BA_+}{A^2}
\left(\frac{BA^2_+}{A^2}\right)_-\right)_-\right]T^3
+
$$
$$
\left\{ 4\frac{BA_+}{A^2}
\left(\frac{BA_+}{A^2}\left(\frac{BA^2_+}{A^2}\right)_-\right)_-
\!\!\!\!\! -
4\left(\frac{BA_+}{A^2}
\left(\frac{BA_+}{A^2}
\left(\frac{BA^2_+}{A^2}\right)_-\right)_-\right)_-
\!\!\!\!\! + \frac{B}{A^2}\left(\frac{BA^2_+}{A^2}\right)_-^2
\!\!\! - \left(\frac{B}{A^2}
 \left(\frac{BA^2_+}{A^2}\right)_-^2\right)_-
\right\}T^4 +
$$
$$
+\ldots
$$
Clearly, all terms before every power
of $T$ gather into ${\cal P}$-positive expression:
$$
F(\tilde T)=A_+T + \left(\frac{BA^2_+}{A^2}\right)_+\!\!\!T^2 -
\,2\!\left[\frac{BA_+}{A^2}\left(\frac{BA^2_+}{A^2}\right)_-\right]_+
\!\!\!T^3 +\left[ 4\frac{BA_+}{A^2}
\left(\frac{BA_+}{A^2}\left(\frac{BA^2_+}{A^2}\right)_-\right)_-
\!\!\!\!\! +
\frac{B}{A^2}\left(\frac{BA^2_+}{A^2}\right)_-^2\right]_+
\!\!\!T^4 +
\ldots
$$
The r.h.s. contains only pure $+$ projections, but note that in this
general situation the  renormalized $F(\tilde T)$ is no longer
quadratic in $T$, as original $F(T)$ was.

\subsubsection{Concrete Example \label{241}}
To the clarify last statement,  consider a simple example.
Let $\Pm$ pick up only terms of exactly the order $\e$,
$\Pm F= \left(\frac{\partial F}{\partial \e}\right)_{\e=0}$,
and let
$$
F(T)=T+b\e T^2+c\e^2 T^3
$$
For such $F(T)$, we can easily check
that there is only one nonzero $Q$, namely $Q(\goo)=-b\e$.
It follows that $\tT=T-b\e T^2$.
Thus renormalized $F(T)$ is no longer quadratic in $T$:
\be
F(T)=F(\tT(T))=T-b\e T^2+b\e (T-b\e T^2)^2+c\e^2 (T-b\e T^2)^3
= \nn \\
=T+(c-2b^2)\e^2T^3+(b^3-3cb)\e^3T^4+3cb^2\e^4T^5-cb^3\e^5T^6
\label{epsilonP}
\ee
Naively this contradicts all the previous examples:
with our previous experience we could rather expect
$F(\tT)=T+c\e^2 T^3$.
However, the point is that $Q$ which would renormalize $F$
in {\it such} a way, would contain ${\cal P}$-positive
terms and thus can not arise from the forest formula.
The forest formula provides $Q(\Gamma)$
which satisfy the constraint (\ref{prob2}).

This is our first example, demonstrating that renormalized $F$
can have very different shape
-- more sophisticated or simplified --
from the original one.
This implies that the second-level forest formula (\ref{fof2}),
adequately describing this shape modification,
is not quite trivial.
Let us show how it reproduces eq.(\ref{epsilonP}).

For $F(\goo)$ we will have the same formula and result as
in~(\ref{fof2_goo}).
For $F(\gooo)$ we have (compare with~(\ref{fof2_gooo})):
\be
F_R(\gooo)=F(\gooo)+\nn\\
+2\left\{F(\goo)\frac{1}{F(\go)}(-\Pm)F(\goo)+
F(\go)\frac{1}{F(\go)}(-\Pm)
\left[F(\goo)\frac{1}{F(\go)}(-\Pm)F(\goo)\right]\right\}=
\nn
\ee
$$
=c\e^2+2\left\{b\e (-\Pm)b\e+(-\Pm)\big[b\e(-\Pm)b\e\big]\right\}
$$
$(-\Pm)\big[b\e(-\Pm)b\e\big]=0$ and we obtain
$F_R(\gooo)=c\e^2-2b^2\e^2$.\
In a similar way we can reproduce coefficients in front of
$T^4,\ T^5,\ T^6$.
The interesting point is to understand, why $F_R(\Gamma)$ vanishes
for chain graphs with more than 6 vertices, as predicted by
(\ref{epsilonP}).
In fact this is also almost obvious:
it is easy to observe that trees with embedded
box-graphs are not allowed, because $\Pm$ in (\ref{fof2}) acts
upwards along the branches of the tree,
and $\Pm \e^n=0$ for $n>1$.
All our boxes can contain only 2 vertices and the only nonzero
$F(\Gamma)$ are those for chain graphs with 1, 2 or 3 vertices.
Multiplication of 2 by 3 gives 6 --
and for all the chain graphs with more than six vertices
$F_R(\Gamma)$ vanishes, in nice agreement with~(\ref{epsilonP}).

\subsubsection{Another example \label{242}}

To further deepen our understanding of problems which are solved
with the help of forest formula,
consider one more example with the
same peculiar projector as in previous section \ref{241}
but a different
$$F(T)=\frac{1}{\e} T+b\e T^2$$
In this case there is a single
non-vanishing counter-term $Q(\Gamma)$ :
$$
\frac{1}{\e} Q(\goo)=-\Pm\left\{F(\goo)\right\}=
-\Pm\big\{b\e\big\}=-b\e
$$
$$
Q(\goo)=-b\e^2
$$
We see that ${\cal P}_+ Q(\goo)=-b\e^2 \neq 0$,
what contradicts the condition (\ref{prob2}).
This is an example, where appropriate modification of (\ref{prob2}),
mentioned in footnote \ref{footnote} is unavoidable:
it is $F(\go)Q(\goo)$ rather than $Q(\goo)$ itself that
should be ${\cal P}$-negative.
This is indeed the case:
 ${\cal P}_+F(\go)Q(\goo)=-{\cal P}_+b\e=0$.

\subsection{Loop diagrams, a need for vacuum energy:
$F(T) \rightarrow F(v,T) = v + T + \epsilon T^2$}

Let us now represent $T$ and $T^2$ by two loop diagrams \loo \ and
\looo. However,  for both these diagrams $\Gamma/\Gamma = \lo$ is
a vertex of valence zero (no external legs), which is not
represented in our function $F(T)$. In order to make formula
(\ref{fof}) applicable in this situation, we need to introduce an
additional term into $F$, associated with this diagram. If we call
the corresponding coupling constant (vacuum energy) $v$, then our
new \be F(v,T) = v+T+\epsilon T^2 = \lo +\loo + \epsilon \cdot
\!\!\looo \ee Now (\ref{fof}) can be applied, and the only
non-vanishing $Q(\Gamma)$ is \be Q\left(\looo\right) = -{\cal P}_-
\left\{ F\left(\looo\right) \right\} = -\epsilon \label{1lp}\ee
Since corrections to the $T$ and $v$ correspond to the graphs with
two and zero external legs respectively, eq.(\ref{1lp}) provides
only a shift of $v$: \be \tilde v = v + Q\left(\looo\right) =
v-\epsilon T^2, \ \ \ \tilde T = T \ee
so that renormalized $F(\tilde v, \tilde
T) = \tilde v + T +\epsilon T^2 = v + T$ is indeed independent of
$\epsilon$.

\subsection{Mixed chain-loop diagrams,
restoration of tensorial structures \label{2.5}}

Let us now represent $T$ by the loop \loo, but leave $T^2$
represented by the chain diagram \goo\ . This time we have two
bare couplings and
\be F(v,T) = \alpha v + (a+b)T + \epsilon c T^2
= \alpha \lo + a\ \go + b\loo + \epsilon c\ \goo
\ee
Now we have
from~(\ref{fof}): \be a Q\left(\goo\right) = -{\cal
P}_-\left\{F\left( \goo\right) \right\} = -\epsilon c \ee \be
\alpha Q\left(\looo\right) = -{\cal P}_-\left\{F\left( \loo\right)
Q\left(\goo\right) \right\} = -b \left(-\frac{\epsilon
c}{\alpha}\right) = \frac{\epsilon bc}{\alpha} \label{chainloop1}
\ee \be a Q\left(\gooo\right) = -{\cal
P}_-\left\{2Q\left(\goo\right) F\left( \goo\right) \right\} =
2\frac{(\epsilon c)^2}{a} \ee \be \alpha Q\left(\loooo\right) =
-{\cal P}_-\left\{ F\left( \loo\right)Q(\gooo) \right\} =
-\frac{b}{\alpha}{\cal P}_-\left\{ Q(\gooo)
\right\}=-2\frac{(\epsilon c)^2}{a}\frac{b}{\alpha}
\label{chainloop2} \ee To understand why the naively expected
coefficients $2$, $3$ -- the numbers of ways we can cut the
corresponding chain graph from the loop one,-- do not appear in
(\ref{chainloop1}) and (\ref{chainloop2}), the tensor structure of
$v$ and $T$ vertices should be restored. After this is done, our
problem acquires the following form:
$$\al \tilde v + a\tT^{ij}+b\tT^{ll} +
\epsilon c \tT^{il}\tT^{lj}=\al v + aT^{ij}+bT^{ll}$$
Due to the presence of two different tensorial structures
this is actually a system of two equations:
$$
a\tT^{ij}+ \epsilon c \tT^{il}\tT^{lj}=aT^{ij}
$$
$$
\al \tilde v +b\tT^{ll}=\al v
$$
Thus $\al \tilde v =\al v - b\tT^{ll}$, and the loop graphs
\looo, \loooo, \ldots \ are just the corresponding
chain graphs \goo, \gooo,~\ldots \ with ends glued together.

For renormalization of $v$ and $T$ we obtain:
\be
\tilde v = \lo +
Q\left(\looo\right)\looo+Q\left(\loooo\right)\loooo + \ldots
= v+ \frac{\epsilon cb}{a\alpha}T^2 -
2\frac{(\epsilon c)^2b}{a\alpha}T^3 + \ldots, \nn \\
\tilde T = \go + Q\left(\goo\right)\goo+Q\left(\gooo\right)\gooo +
\ldots = T -\frac{\epsilon c}{a} T^2 + 2\left(\frac{\epsilon
c}{a}\right)^2T^3 - \ldots
\ee
Since the values of $Q[n^{\mbox{\small\small (loop)}}]$
for the loop diagrams are determined
by the values of $Q[n]$ for the corresponding chain diagrams,
we can perform summation with the help of a generating function,
in the same way as it was done in (\ref{summQ}):
$$
F(\tilde v,\tilde T) = \alpha \tilde v + (a+b)\tilde T +
\epsilon c\tilde T^2 =
\alpha\left(v+ \frac{\epsilon bc}{a\alpha}T^2 + \ldots\right)
+ (a+b)\left(T -\frac{\epsilon c}{a} T^2 + \ldots\right) +
\epsilon c \left(T -\frac{\epsilon c}{a} T^2 + \ldots\right)^2 =
$$ $$
=\alpha v + (a+b)T
$$
Note, that even if $\alpha$ was originally put to zero, it
acquires $\alpha$-independent correction in renormalization
process.
The zero-valence vertex is automatically generated by the
forest formula.
However this happens in a singular way:
the $\al$-dependent correction is first divided
and then multiplied by $\al \rightarrow 0$.

\section{Generic polynomial from the valence-two vertex}

Now we are ready to attack a slightly more general problem than in
the previous sections:
take for a function $F(T)$ a sum of two general
polynomials with the same projector as in (\ref{trivproj}):
$$
F(T)=R(T)+\epsilon G(T)
$$
Renormalization problem reduces to the following equation:
$$
R(\tT)+\e G(\tT)=R(T),
$$
and solution is represented as power series in $\e$:
\be
\tT=T-\e\frac{G(T)}{R'(T)}+\e^2\frac{G(T)}{2R^{'3}(T)}(2R'(T)G'(T)-
R''(T)G(T))+\ldots
\label{genpol}
\ee
Immediate question is what is the origin of $R'$ in denominator
from the point of view of the forest formula.

\subsection{The simplest nontrivial $F(T)=T+\al T^2+\e T^3$}

We begin from example with the simplest nontrivial $R$:
$R(T)=T+\al T^2$ and $G(T)=\e T^3$.
Substituting this into (\ref{genpol}), we get:
\be \tT=T-\e \frac{T^3}{1+2\al T}+\e^2\frac{T^5(3+5\al T)}{(1+2\al
T)^3}+\ldots \label{genpolt} \ee
and expanding denominators in $\al T$, obtain:
\be \tT=T-\e (T^3-2\al T^4+4\al^2 T^5-8\al^3
T^6+\ldots)+\e^2(3T^5-13\al T^6+42\al^2 T^7+\ldots )+\ldots
\label{genpoltexp} \ee

Now we attack this problem with the forest formula:
$$
F(T)=T+\al T^2+\e T^3=\go+\al\ \goo+\e\ \gooo
$$ $$
Q[3]=Q(\gooo)=-\e,
$$ $$
Q[4]=- \Pm (2F[2]Q[3])=2\al\e,
$$ $$
Q[5]=-\Pm(2F[2]Q[4]+3F[3]Q[3])=3\e^2-4\al^2\e,
$$ $$
Q[6]=-\Pm(F[2]Q^2[3]+2F[2]Q[5]+3F[3]Q[4])=8\al^3\e-13\al\e^2
$$
Thus for $\tT$ we have:
$$
\tT=T-\e T^3+2\al\e
T^4+(3\e^2-4\al^2\e)T^5+(8\al^3\e-13\al\e^2)T^6+\ldots=
$$
$$
=T-\e(T^3-2\al T^4+4\al^2 T^5-8\al^3 T^6+\ldots)+\e^2(3T^5-13\al
T^6+\ldots)+\ldots
$$
This is exactly the same as (\ref{genpoltexp}), but  expanded
first in $\al T$ and then in $\e$.

It is instructive to sum explicitly all the terms,
contributing to the first and second order in $\e$.
As usual, this can be done with the help of generating functions
$Q_{(1)}(t)=\sum_{n=3}t^nQ_{(1)}[n]$ and
$Q_{(2)}(t)=\sum_{n=3}t^nQ_{(2)}[n]$, where $Q_{(1)}[n]$
and $Q_{(2)}[n]$ denote the contributions to in $Q[n]$
of orders $\e$ and $\e^2$ respectively.

Since $Q[n]$ does not contain terms of the order $\e^0$,
only the terms with a single box-subgraph contribute to
$Q_{(1)}[n+1]$:
$$
Q_{(1)}[n+1]=-\Pm(2F[2]Q_{(1)}[n])
$$
For $Q_{(1)}(t)$ this yields
$$
Q_{(1)}(t)=Q[3]t^3-2\al t Q_{(1)}(t), \ \
Q_{(1)}(t)=\frac{t^3Q[3]}{1+2\al t}=-\e\frac{t^3}{1+2\al t}
$$
For $Q_{(2)}[n+1]$ we should take
three contributions into account:
$F[2]Q_{(2)}[n]$, $F[3]Q_{(1)}[n]$ and
$F[2]Q_{(1)}[n]Q_{(1)}[k]$:
$$
Q_{(2)}[n+3]=-\Pm\left(2F[2]Q_{(2)}[n+2]+3F[3]Q_{(1)}[n+1]+
F[2]\sum_{k=3}^{n}Q_{(1)}[k]Q_{(1)}[n+3-k]\right)
$$
Therefore $Q_{(2)}(t)$ satisfies
$$
Q_{(2)}=-\left(2tF[2]Q_{(2)}+3t^2F[3]Q_{(1)}+F[2]Q_{(1)}^2\right)
$$
and
$$
Q_{(2)}=-\frac{3\e t^2 Q^{(1)}+\al Q^{(1) 2} }{1+2\al
t}=\frac{\e^2t^5(3+5\al t)}{(1+2\al t)^3}
$$
Thus, after re-summation, we reproduce (\ref{genpolt}).

\subsection{General $F(T)$}

To understand the way, in which the forest formula can
differentiate polynomials and provide $R'(T)$ in denominator
in (\ref{genpol})),
it is sufficient to sum up the terms of the order~$\e^1$.
Exactly as in previous subsection, this is
by solving an equation for the generating function
$$
F(T)=T+\sum_{n=2}^{R}r_nT^n+\e\left(\sum_{n=1}^{G}g_nT^n\right),
$$
Here $R$ and $G$ denote degrees of the polynomials $R(T)$ and
$G(T)$ respectively, and
for the sake of simplicity $r_1$ was put to one.
If
$$
Q_{(1)}(t)=\sum_{n=1}t^nQ_{(1)}[n],
$$
then for $n\geq R$:
$$
Q_{(1)}[n]=-\Pm^{(1)}\left\{
2F[2]Q_{(1)}[n-1]+3F[3]Q_{(1)}[n-2]+
4F[4]Q_{(1)}[n-3]+\ldots+RF[R]Q_{(1)}[n-R+1]
\right\}
$$
where $\Pm^{(1)}$ is a projector which pics up
only the $\e^1$-terms.
The coefficients $2$, $3$, \ldots $R$ count
the numbers of ways that we have to cut a box subgraph
of corresponding the length from the chain graph $[n]$.
When expressed in terms of $Q_{(1)}(t)$,
this recursion relation becomes:
$$
Q_{(1)}(t)=-\left(2r_2t+3r_3t^2+\ldots
+Rr_Rt^{R-1}\right)Q_{(1)}(t)
-\left(g_1t-g_2t^2-\ldots-g_Gt^G\right),
$$
and finally
$$
Q_{(1)}(t)=-\frac{G(t)}{R'(t)}
$$
In a similar way we can perform term-by-term re-summations
for all orders in $\e$.

\section{Cubic equation. Valence-3 vertex and loops}

Let now $T$ represent a valence-$3$ vertex.
Then we can use the
forest formula to solve cubic equation
$\tilde T+ \epsilon T^3=T$,
assuming that the projector is again given by (\ref{trivproj}).
Solution is
$$
\tilde T=T-\epsilon T^3+3\epsilon^2 T^5-12\epsilon^3
T^7+55\epsilon^4 T^9+\ldots=T+\sum_{n=1} (-)^n c_n \epsilon^n
T^{2n+1}
$$
where $c_n$ are the numbers of rooted $3$-valent trees
with exactly $n$ vertices, e.g.
\be c_1=1= \ \rtro \ \ \ \ c_2=3=\ \ \rtrou + \ \ \rtrod
+\ \ \rtrom\label{roottree}
\ee
We now reproduce this solution from the forest formula.
Like in the case of valence $2$, there are different
possibilities to represent our $F(T)$ and thus different
realizations of renormalization procedure.

\subsection{First realization}

The first way is to represent $T+\epsilon T^3$ as
\be
F(T) = T +\epsilon T^3 = \tro + \epsilon \ \troo
\ee

\bigskip \bigskip

\noindent
When we start calculating $Q(\Gamma)$,
the forest formula itself
selects the relevant graphs from
the huge set of all possible graphs with
$4$-valence vertices.
$$
\begin{array}{c}
Q\left(\tro\right)=
-{\cal P}_-\left\{F\left(\tro\right) \right\}=0 \\
\\
Q\left(\troo\right)=
-{\cal P}_-\left\{F\left(\troo\right) \right\}=-\epsilon \\
\\
Q\left(\trood\right)=Q\left(\troou\right)=Q\left(\troom\right)=
-{\cal P}_-\left\{F\left(\troo\right)Q\left(\troo\right) \right\}
=\epsilon^2 \\
\\
Q\left(\trooud\right)=\ldots=
-{\cal P}_-\left\{F\left(\troo\right)Q\left(\troo\right)
Q\left(\troo\right) \right\}=-\epsilon^3 \\
\end{array}
$$
It is easy to notice, that since embedded graphs are not allowed,
all non-vanishing $Q(\Gamma)$ will have the form
$$Q(\Gamma)=(-)^{\frac{k-1}{2}}\epsilon^{\frac{k-1}{2}},$$ where $k$
is the number of vertex in $\Gamma$.
Thus the problem reduces to
counting number of graphs in order $k$,
which give nonzero contribution.
However this becomes an easy task if we observe a
one-to-one correspondence between 'triangle' graphs and rooted
trees from (\ref{roottree}):
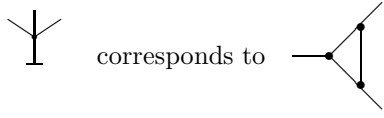
$$
\rtro \ \ \mbox{corresponds to }\ \  \troo \
$$
When we grow three links on the some vertex of original tree,
this corresponds to insertion of \troo \ \ into the
corresponding vertex of our graph, for example:
$$
 \rtrou \ \ \ \longleftrightarrow\ \ \ \troom
$$

\bigskip \bigskip

\noindent
Therefore we obtain  from the forest formula the following
result for $\tilde T$:
$$
\tilde T= T+ Q\left(\troo\right)T^3 +\left\{
Q\left(\trood\right)+Q\left(\troou\right)+
Q\left(\troom\right)\right\}T^5+
$$
$$
+\left\{Q\left(\trooud\right)+\ldots \right\}
T^7+\ldots =\
T + \sum_{n=1} (-)^n c_n \epsilon^n T^{2n+1}
$$

\subsection{Second realization}

The same function $F(T)$ can be represented in a different way:
\be
F(T)=T + \epsilon T^3 = \lcho + 3\epsilon \ \lchoo
\ee
This time we put additional coefficient $3$, reflecting
the symmetry of the problem. It becomes especially natural if we
restore indices of a triple totally-symmetric vertex: $T
\rightarrow T_{ijk}$, so that (sums implied over repeated indices)
\be F_{ijk}(T) = T_{ijk} + \epsilon \Big((TT)_{im}T_{mjk} +
(TT)_{jm}T_{imk} + (TT)_{km}T_{ijm}\Big), \ \ \ \ (TT)_{im} =
T_{ipq}T_{mpq} \ee and actually there are three different one-loop
diagrams.

The first counter-terms are: \be Q\left(\lchoo\right) = -{\cal
P}_-\left\{ F\left(\lchoo\right) \right\} = -\epsilon \ee Note
that there is exactly the same diagram at the r.h.s. and the
l.h.s., no tripling occurs in this formula.

At the next level there are diagrams of three different
topologies:
\be
Q\left(\lchooo\right) = -3{\cal P}_-\left\{
F\left(\lchoo\right) Q\left(\lchoo\right) \right\} =
3\epsilon^2,\label{topol1}
\ee

\be
Q\left(\lcholo\right) = -2{\cal P}_-\left\{
F\left(\lchooo\right) Q\left(\lchooo\right) \right\} =
2\epsilon^2,\label{topol2}
\ee

\be
Q\left(\lchllu \right)=Q\left(\lchlld \right) = -2{\cal
P}_-\left\{ F\left(\lchoo\right) Q\left(\lchoo\right) \right\} =
2\epsilon^2\label{topol3},
\ee
where coefficients $3$ in (\ref{topol1}) and $2$ in
(\ref{topol2}) and (\ref{topol3}) correspond to the number of
possible choices of box-subgraphs.

Altogether, taking into account three
different orientations of the diagrams,
we get for the counter-term $Q(T)$:
$$
\tilde T = T - 3\epsilon T^3 +
3\cdot (3+2+2\times 2)\epsilon^2 T^5 + \ldots
$$
and
$$
F(\tilde T) = \tilde T + 3\epsilon \tilde T^3 =
\Big(T - 3\epsilon T^3 + 27 \epsilon^2 T^5 + \ldots\Big) +
3\epsilon \Big(T - 3\epsilon T^3 + 27 \epsilon^2 T^5 +
\ldots\Big)^3 = T
$$

\section{Cubic equation. Two vertices: of valence 2 and 3}

We already encountered a problem with more than one
coupling constant, but this was a rather trivial example.
Now we reveal some more peculiarities of the forest formula's
work with several couplings.
Consider
$$
F(T)=T_2+\e_2 T_3^2+ T_3+ \e_3T_3^3+3\e_3^2T_3^3T_2=
\vspace{-0.5cm}
$$
\be
=\ \go+\e_2 \buble+ \tro+ \e_3\ \troo + \e_3^2 \
\troodotc+ \e_3^2\ \troodota + \e_3^2\ \troodotb
\ee

\bigskip

and $\Pm$ pics up all the terms $\e_2^{\alpha} \e_3^{\beta}$
with all $|\alpha|+|\beta| \geq 1$.

Again, like section \ref{2.5}, due to the presence of different
tensorial structures, eq. $F(\tilde T) = T_2+T_3$ implies
renormalization of two distinct functions:
$$
\tT_2+\e_2 \tT_3^2=T_2
$$
$$
\tT_3+ \e_3\tT_3^3+3\e_3^2\tT_3^3\tT_2=T_3
$$
However this time these two equations are interrelated,
and iteration procedure should determine $\tT_2$ and $\tT_3$
simultaneously. Let us try to solve this system with the help
of the forest formula.
The first graph is simple:
\be
Q\left(\buble\right)=-\Pm\left\{F\left(\buble\right)\right\}
=-\e_2 \label{buble}
\ee
However, when we look at the next graph with two external legs,
we encounter a difficulty:
$$
Q\left(\bublel\right)=
-\Pm\left\{2F\left(\buble\right)Q\left(\troo\right)\right\}= \ ?
$$
It turns out that we should first calculate
\be
Q\left(\troo\right)=-\Pm\left\{F\left(\troo\right) \right\}
=-\e_3 \label{triangle}
\ee
and only after that return to
\be
Q\left(\bublel\right)=
-\Pm\left\{2F\left(\buble\right)Q\left(\troo\right)\right\}=
2\e_2\e_3 \label{33}
\ee
The next graphs are:
\be
Q\left(\troodotc\right)=-\Pm\left\{F\left(\troodotc\right)\right\}
=-\e_3^2 \ \ (\times 3), \label{triangledot}\\
\nonumber\\
Q\left(\troom\right)=-\Pm\left\{F\left(\troo\right)
Q\left(\troo\right)\right\}=\e_3^2 \ \ (\times 3),\label{36}\\
\nonumber\\
Q\left(\troodloop\right)=-\Pm\left\{F\left(\troodotc\right)
Q\left(\buble\right)\right\}=\e^2_3\e_2 \ \ (\times 3) \label{37} \\
\nonumber\\
Q\left(\bubleldot\right)=-\Pm\left\{F\left(\buble\right)
Q\left(\troodota\right)\right\}=\e_2\e_3^2
\ \ (\times 4) \label{38} \\
\nonumber\\
Q\left(\bublelmidledot\right)=-\Pm\left\{2F\left(\buble\right)
Q\left(\troodotc\right)\right\}=2\e_2\e_3^2
\ \ (\times 1) \label{39}
\ee
where $(\times 3), (\times 4), \ldots $ denote the numbers
of possible positions of the dot or,
in the case of (\ref{36}), the number of orientations
of the corresponding graph.
Again note, that for  (\ref{37}) we need  (\ref{buble}), for
(\ref{38}) and (\ref{39}) we need (\ref{triangledot}),
although these are graphs with different number of external legs.

The forest formula selects by itself the graphs which contribute to
renormalization.
It is because of connection between
renormalization processes for $T_2$ and $T_3$ that
we cannot determine first $Q$ for graphs with two external legs,
and then for those with
three external legs -- this should be done simultaneously,
order by order:
\be
\tT_2=T_2-\e_2
\stackrel{\mbox{\tiny{(\ref{buble})}}}
{T_3^2}+2\e_2\e_3\stackrel{\mbox{\tiny{(\ref{33})}}}{T_3^4}+
4\e_2\e_3^2\stackrel{\mbox{\tiny{(\ref{38})}}}
{T_2T_3^4}+2\e_2\e_3^2\stackrel{\mbox{\tiny{(\ref{39})}}}
{T_2T_3^4}+\ldots
\label{t2} \ee \be
\tT_3=T_3-\e_3\stackrel{\mbox{\tiny{(\ref{triangle})}}}
{T_3^3}-3\e_3^2\stackrel{\mbox{\tiny{(\ref{triangledot})}}}
{T_2T_3^3}+
3\e_3^2\stackrel{\mbox{\tiny{(\ref{36})}}}
{T_3^5}+3\e_2\e_3^2\stackrel{\mbox{\tiny{(\ref{37})}}}
{T_3^5}+\ldots
\label{t3}
\ee
To make clear the origin of every term, we keep reference to
the graph, which it originates from.
Now it is a simple exercise, to check that
(\ref{t2}) \& (\ref{t3}) solve
original problem.

\part{QFT examples}

QFT is a rather natural place to apply forest formulas --
and it is what they were originally designed for.
Perturbation theories in QFT often suffer from divergences,
originating from momentum integrations in loop diagrams,
i.e. from sums over tensorial indices in our diagram
expansions.
To handle divergent integral one applies
{\it regularizations}, which introduce dependence
on additional parameters,
like an ultraviolet cut-off $\Lambda$ in naive schemes,
or $\epsilon = 4-d$ in dimensional regularization, or
masses $M_i$ in the Pauli-Villars approach.
At the end of the calculation these parameters need be
eliminated.
This can {\it not} be achieved by simply throwing them away
from the answer, because this would change
mutual relations between different correlators, i.e.
the shape of partition function.
The risk is that after an arbitrary change we can loose
possibility to obtain this partition function
and these correlators from a single functional integral:
they can fail to represent any kind of quantum mechanical
system.
Such risks are eliminated if one does not change the
shape of partition function, but modifies its arguments, i.e.
the {\it bare} couplings, which are exactly our parameters
$\tT$ from the previous sections.
Forest formula gives us an expression for these bare couplings
$\bar T$ in terms of the physical constants $T$:
$$\tT \equiv T+Q(T),$$
where $Q(T)$ are singular counter-terms,
which make partition function finite.
In this sense the forest formula provides the resolution
of renormalization problem in QFT.

Forest formula operation does not depend on the nature of
the function $F(T)$ which we want to renormalize.
The only thing one should care about to perform a
self-consistent renormalization is the validity of
the {\it vertex criterium}:
for every graph $\Gamma$ with non-vanishing $Q(\Gamma)$
there should be a vertex $\Gamma/\Gamma$ in $F(T)$, i.e.
$F(\Gamma/\Gamma)\neq 0$ if $Q(\Gamma)\neq 0$.
If this is true, renormalization in QFT
is performed at the level of partition function,
i.e. for all the correlators at once.
However it is often possible to restrict
consideration to correlators with particular external
structure.

\section{$\phi^4$ in one and two loops \label{sec6}}

The simplest non-trivial example is provided by renormalization
of the $4$-point function in the $\phi^4$ theory at two loops
(i.e. in the order $T^3$).
This example is still rather simple, nevertheless, it dmonstrates
the main features and peculiarities of the forest formula.

Our theory is:
\be
L(\phi)=\frac{1}{2}(\partial_{\mu}\phi)^2-
\frac{1}{2}m^2\phi^2-\frac{T}{4!}\phi^4
\label{phi4}
\ee
We assume, that renormalization of field and of the mass term
has already been performed, and thus
we do not consider renormalization of kinetic and mass terms:
valence-$2$ vertices and divergent graphs with 2 external lines
are not considered
(see, however, sec. \textbf{\ref{app}} in the Appendix below).
We begin directly with the 4 point function,
which is written as a sum of Feynman diagrams:
\be
F(T)=\oloop + \uloop+\tloop+\sloop+
\left(\siiloop+\ldots\right)+ \left(\siirloop+\ldots\right)+
\left(\siilloop+\ldots\right) \label{4point}
\ee
Triple dots stand for omitted diagrams in the $u$ and $s$ channels.
For
$Z(\Gamma|T)$ we take the following functions:
$$
Z\left(\oloop|T\right)=-iT,\ \ \
Z\left(\mbox{1-loop diagramm}|T\right)=(-iT)^2, \ \ \
Z\left(\mbox{2-loop diagramm}|T\right)=(-iT)^3, \ \ldots
$$
Then the lowest coefficients $\hat F(\Gamma)$ are:
$$
F\left(\oloop\right)=1,
$$
$$
F^{(1,t)}(p^2) \equiv F\left(\sloop\right)
=\frac{1}{2}\int\frac{d^4k}{(2\pi)^4}
\frac{i}{k^2-m^2}\frac{i}{(k+p)^2-m^2}
$$
Pauli-Villars regularization substitutes $F^{(1,t)}(p^2)$ by
\be
F^{(1,t)}_{reg\ M}(p^2)=\frac{1}{2}\int\frac{d^4k}{(2\pi)^4}
\left(\frac{i}{k^2-m^2}-\frac{i}{k^2-M^2}\right)
\left(\frac{i}{(k+p)^2-m^2}-\frac{i}{(k+p)^2-M^2}\right)=\nn\\
=\frac{i}{32\pi^2}
\left\{-\log\frac{M^2}{\mu^2}+\int_0^1dx
\log\frac{m^2-x(1-x)p^2}{\mu^2}+O(\frac{1}{M})\right\}=
\frac{i}{32\pi^2}
\left\{-\log\frac{M^2}{\mu^2}+I^{(1,t)}_{fin}(p^2) \right\}
\label{F1tM}
\ee
Similar formulas can be written for
$F^{(1,s)}_{reg\ M}$ and $F^{(1,u)}_{reg\ M}$,
in all of them $p^2$ denote the squared momentum in the loop.
Evaluation of $F^{(2,t^2)}(p^2) \equiv F\left(\siiloop\right)$
is straightforward, it is just the square of $F^{(1,s)}$:
\be
F^{(2,t^2)}_{reg\ M}(p^2)=
\left(\frac{i}{32\pi^2}\right)^2\left\{\left(\lnM\right)^2
-2\lnM I^{(1,t)}_{fin}(p^2)+
\left(I^{(1,t)}_{fin}(p^2)\right)^2\right\}
\ee
Another two-loop diagram is more complicated,
moreover, it contains overlapping divergencies:
$$
F^{(2,t\cdot(s+u))}(p,p_4) \equiv
F\left(\, \siirloop\right)=\int\frac{d^4q}{(2\pi)^4}
\frac{i}{q^2-m^2}\frac{i}{(p-q)^2-m^2} F^{(1,t)}((q-p_4)^2)
$$
Here $q$ is the momentum, running through the
bigger loop, $p$ is the sum of momenta in the bigger loop???
and  $p_4$ is the momentum of
the right bottom external leg of the diagram.
Substituting $F^{(1,t)}$ from (\ref{F1tM}) we obtain:
\be
F^{(2,t\cdot(s+u))}_{reg\ M}(p,p_4) =-2\frac{i}{32\pi^2}\lnM
F^{(1,t)}_{reg\ M}(p^2)+\nn\\
+\frac{i}{32\pi^2}\int\frac{d^4q}{(2\pi)^4}
\left(\frac{i}{q^2-m^2}-\frac{i}{q^2-M^2}\right)
\left(\frac{i}{(q-p)^2-m^2}-\frac{i}{(q-p)^2-M^2}\right)\times
I^{(1,s)}_{fin}((q-p_4)^2)=\\\nn
=\left(\frac{i}{32\pi^2}\right)^2\left\{2\left(\lnM\right)^2-2\lnM
I^{(1,s)}_{fin}(p^2)-\left(\lnM\right)^2+
I^{(2,t\cdot(s+u))}_{fin}(p,p_4)\right\}
\ee

Since expression for every diagram naturally
decomposes into two parts,
one finite in the limit when $M\longrightarrow\infty$,
and another, divergent, behaves like $\left(\lnM\right)^n$ with
some positive integer $n$,
we ask ${\cal P}_-$ to pick up the contributions like
$\left(\lnM\right)^n$ with $n>0$.

Now we can apply the forest formula (\ref{fof}):

\be
Q\left(\sloop\right)=Q\left(\ \ \tloop\ \right)=
Q\left(\ \uloop\right)=
-\Pm\left\{F\left(\sloop\right)\right\}=
\frac{i}{32\pi^2}\lnM
\ee
\be
Q\left(\siiloop\ \, \right)=
-\Pm\left\{F\left(\siiloop\ \, \right)+
2F\left(\sloop\right)Q\left(\sloop\right)\right\}=
\ee
$$
=-\left[
\left(\frac{i}{32\pi^2}\right)^2\left\{\left(\lnM\right)^2-
2\lnM I^{(1,s)}_{fin}(p^2)\right\}+
2\times \frac{i}{32\pi^2}
\left\{-\log\frac{M^2}{\mu^2}+I^{(1,t)}_{fin}(p^2) \right\}
\left\{\frac{i}{32\pi^2}\lnM\right\}
\right]=
$$
\be
=\left(\frac{i}{32\pi^2}\right)^2\left(\lnM\right)^2
\ee

$$
Q\left(\siirloop\right)=
-\Pm\left(F\left(\siirloop\right)+
F\left(\sloop\right)Q\left(\ \tloop\ \right)+
F\left(\sloop\right)Q\left(\,\uloop\right)\right)=
$$
$$
=-\left(\frac{i}{32\pi^2}\right)^2
\Pm\left[\left(\lnM\right)^2-2\lnM
I^{(1,s)}_{fin}(p^2)+
I^{(2,t\cdot(s+u))}_{fin}(p^2)-
2\left\{-\lnM+I^{(1,t)}_{fin}(p^2)\right\}\lnM
\right]=
$$
\be
=-3\left(\frac{i}{32\pi^2}\right)^2\left(\lnM\right)^2
\ee
Collecting all these $Q$'s into a single formula for $\tT$,
we obtain:
$$-i\tT=-iT+3\frac{i}{32\pi^2}\lnM(-iT)^2-3
\left(\frac{i}{32\pi^2}\right)^2\left(\lnM\right)^2(-iT)^3$$
$$\tT=T+\frac{3}{32\pi^2}\lnM T^2-
\frac{3}{(32\pi^2)^2}\left(\lnM\right)^2T^3$$
Such $\tT$ indeed makes $F(T)$ $\cal{P}$-positive, and again,
the resulting $F(\tilde T)$ can be most straightforwardly
obtained from the second-level forest formula (\ref{fof2}):
$$
F_R(T)=F(\tT)
$$
$$F_R\left(\oloop\right)=F\left(\oloop\right)=1$$
$$
F_R\left(\sloop\right)=F\left(\sloop\right)+
F\left(\oloop\right)\frac{1}{F\left(\oloop\right)}
\mPm F\left(\sloop\right)=\Pp F\left(\sloop\right)=$$
$$
=\frac{i}{32\pi^2}\int_0^1dx\log\frac{m^2-x(1-x)p^2}{\mu^2}
$$
At the level of double-loop graphs we have to sum over the
following forests for \siiloop:

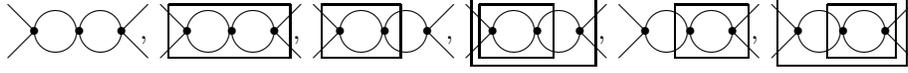
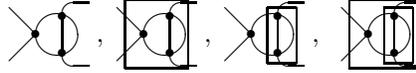
\bigskip

$$\siiloop, \ \siiloopboxed, \
\siilooplboxed, \ \siiloopalboxed, \ \siilooprboxed,
\ \siilooparboxed
$$

\bigskip

and  for \siirloop:

$$
\siirloop, \  \siirloopboxed, \  \siirlooprboxed, \ \
\siirlooparboxed
$$

\bigskip

\noindent
According to the rules described in section \textbf{\ref{intro}},
the forest \siiloopboxed \ \  corresponds to
$$\mPm F\left(\siiloop\ \right)$$
(from now on we omit $F\left(\oloop\right)=1$),
while  \siiloopalboxed \ \ corresponds to
$$\mPm \left\{F\left(\sloop\ \right)
\mPm F\left(\sloop\ \right)\right\}$$
If we represent the action of $\Pm$ on the graph
by surrounding it with an oval
and denote the vertex which substitutes the contracted graph
through $\contracted$, we can symbolically
write expression for the $R(\Gamma)\equiv F_R(\Gamma)$ as:
\be
R\left(\sloop\right)=\sloop - \sloopovaled \label{roper1}
\ee
$$
R\left(\siiloop\ \right)=
\siiloop\ \ -\ \ \siiloopovaled\ \ -
\ \ \sloopleftdot \times \sloopovaled \ \ +
\ \ \sloopleftdotovaled \times \sloopovaled \ \ -
$$
\be
-\ \ \slooprightdot \times \sloopovaled \ \
+\ \ \slooprightdotovaled \times \sloopovaled \ \
\label{roper2}
\ee

\bigskip

\noindent
Now it is easy to recognize in $R(\Gamma)\equiv F_R(\Gamma)$
the Bogolubov's {\it $R$-operation} \cite{Bogo,Zav}
and its graphical representation in
eqs.(\ref{roper1}) and (\ref{roper2}).

\section{Renormalizable and non-renormalizable QFT example:
$\phi^4$ in $d=4$ and $d\geq5$}

In this section we briefly illustrate the difference between
renormalizable and non-renormalizable QFT models.
Note, that the forest formula
works order-by-order in coupling constant,
thus all graphs of a given order should be included.
Coming back to eq.(\ref{4point}) for the $\phi^4$ theory,
we observe that in the third order there are more diagrams,
for example,
\be\sixloop\label{sixloop}\ee
It was omitted in eq.(\ref{4point}), because for $d=4$ its
Feynman amplitude does not contain divergent terms:
\be
\Pm\left\{F\left(\sixloop\right)\right\}=0 \label{4dimsixloop}
\ee
However, this statement depends on space-time dimension $d$, and
this diagram diverges as $M^{d-6}$ for $d\geq 6$.
For $d=5$ this graph is finite, however, the diagram
\be
\sixIbuble
\label{sixI}
\ee
does diverge.
This means, that in order to satisfy the
{\it vertex criterium} and perform a self-consistent
renormalization we have to
include an elementary six-valence vertex $T_6$
into $F(T)$, in addition to the elementary $4$-vertex $T_4$:
$$
F\left(\six\right)Q\left(\sixloop\right)=
-\Pm\left\{F\left(\sixloop\right)\right\}\neq 0
\ \Longrightarrow \ F\left(\six\right)\neq 0,
$$
After this the contributions from other graphs
should be added, until all
diagrams up to the third order will be taken into account.
The diagrams to be added are, for example:
 $$\sixbuble, \ \ \eightbuble$$
Both of them diverge, since they contain a loop with
two propagators, and they give contribution  to $Q(\Gamma)$.
The first one contributes additionally to
renormalization of $T_6$, however the second
diagram  generates $T_8$ -- an elementary eight-valence vertex:
$$
F\left(\eight\right)Q\left(\eightbuble\right)=
-\Pm\left\{F\left(\eightbuble\right)\right\}\neq 0
\ \Longrightarrow \ F\left(\eight\right)\neq 0,
$$
In turn $T_8$ generates $T_{10}$, and so on and so force.
We see, that for $d \geq 5$,
despite renormalization of $F(T)$ is possible,
an \textit{infinite number} of
vertices should be introduced to perform it self-consistently
even in the third order in $T$.
Such theories are called \textit{non-renormalizable}.

Situation in $d=4$ advantageously differs from higher dimensions.
Due to (\ref{4dimsixloop}), the diagram (\ref{sixloop})
does not contribute to renormalization of $F(T)$ in (\ref{4point})
and generates no new vertices.
And this property is not bounded to the order $T^3$.
It can be shown \cite{Bogo}-\cite{Peskin},  that all
divergencies in graphs with more than $4$ external legs come from
the sub-divergent $4$-valence graphs.
For example, although for $d=4$
$$\Pm\left\{F\left(\sixloopbuble\right)\right\}\neq0,$$
this divergence does not force one to introduce the
bare $T_6$ vertex,
because it is exactly canceled in the forest formula:
$$ F\left(\six\right)Q\left(\sixloop\right)=
\Pm\left\{F\left(\sixloopbuble\right)-
F\left(\sixloop\right)Q\left(\ \tloop\right)\right\}=0,$$
and we can consistently keep $F\left(\six\right)=0$.
In the language of renormalized perturbation theory
this means that all divergencies can be absorbed into a single
counter-term for the $4$-valence vertex:
$Q(\Gamma)$ which makes $F(T)$ or any of the correlators in the
theory finite,  takes non-zero value only
on the graphs with 4  external legs.
Theories, possessing such property, are called
\textit{renormalizable}.

\part{Appendices}

\section{Proof of the forest formula}

After a number of concrete examples that were examined in the
main part of this article, it should be a simple task
to follow the proof of the forest formula.
We remind that for a given function $F(T)$ and linear projector
${\cal P}_-$ it solves the equation (\ref{prob1}),
\be
{\cal P}_-\left\{ F\Big(T + Q(T)\Big)\right\} = 0
\label{prob1_1}
\ee
for the shift $T \rightarrow \tilde T = T + Q(T)$.
In other words, it allows to modify the value of the function
$F$ in desired way, by shifting its {\it argument}
while preserving its {\it shape}.
This is important in physical applications, because one
can cure problems of partition function (e.g. divergences),
without breaking its distinguished properties
(say, keeping it inside the narrow class of integrable
$\tau$-functions).

Representing
$$
F(T+Q(T))=\sum_{\Gamma} \hat F(\Gamma) Z(\Gamma|T+Q(T))
=\sum_{\Gamma} \hat F_R(\Gamma) Z(\Gamma|T)
$$
we see, that since the functions $Z(\Gamma|T)$ form a basis
in the space of all functions of T,
eq.(\ref{prob1_1}) can be satisfied if
and only if $\Pm F_R(\Gamma)=0$ for all
$\Gamma$.
(We remind that projector $\Pm$ does not acts on the $Z(\Gamma|T)$).
To find the coefficients $\hat F_R(\Gamma)$,
we substitute $Q(T)=\sum_{\gamma} \hat Q(\gamma) Z(\gamma|T)$
and compare the two sides of equality:
$$
\sum_{\Gamma} \hat F\big(\Gamma\big) Z
\Big(\Gamma\Big|\ T+\sum_{\gamma} \hat Q(\gamma) Z(\gamma|T)\Big)
=\sum_{\Gamma} \hat F_R(\Gamma) Z(\Gamma|T)
$$
Since
$$Z\Big(\Gamma\Big|\ T+\sum_{\gamma} \hat Q(\gamma) Z(\gamma|T)\Big)=
Z(\gamma|T)+
\sum_{\Up,\gamma: \Up/\gamma=\Gamma} \hat Q(\gamma)
Z(\Up|T)+\sum_{{\Up,\gamma_1,\gamma_2:}{ \Up/(\gamma_1\cdot\gamma_2})
=\Gamma} \hat Q(\gamma_1)\hat Q(\gamma_2)Z(\Up|T)+\ldots$$
we get for $\hat F_R$:
\be
\hat F_R(\Gamma)=\hat F(\Gamma)+
\sum_{\big\{\gamma_1 \cup \ldots \cup \gamma_k\big\}}
\hat F\Big(\Gamma/\gamma_1\ldots \gamma_k\Big)
\hat Q(\gamma_1)\ldots\hat Q(\gamma_k),
\label{F_R}
\ee
$$
\Pm \hat F_R(\Gamma)=0
$$
$$
\Pm\left\{\hat F(\Gamma)+\hat F(\Gamma/\Gamma)\hat Q(\Gamma)+
\mathop{{\sum}'}_{\big\{\gamma_1 \cup \ldots \cup \gamma_k\big\}}
\hat F\Big(\Gamma/\gamma_1\ldots \gamma_k\Big)
\hat Q(\gamma_1)\ldots\hat Q(\gamma_k)\right\}=0,
$$
where prime in $\sum'$ signals that we extracted
the term with $\gamma=\Gamma$ from the sum.
Now, supposing that
${\cal P}_+ \hat F(\Gamma/\Gamma)\hat Q(\Gamma)=0$,
we arrive to the forest formula (\ref{fof}):
$$
\hat F(\Gamma/\Gamma) \hat Q(\Gamma) = -{\cal P}_-\left\{
\hat F(\Gamma) +
\sum_{\big\{\gamma_1 \cup \ldots \cup \gamma_k\big\}}
\hat F\Big(\Gamma/\gamma_1\ldots \gamma_k\Big)
\hat Q(\gamma_1)\ldots\hat Q(\gamma_k)\right\}
$$
Q.E.D.

The second-level forest formula (\ref{fof2})
can be obtained by solving recursion for $\hat Q(\Gamma)$
and substituting the result into~(\ref{F_R}).

\section{Kinetic terms, 1PI diagrams and other peculiarities \label{app}}
Here we briefly comment on some peculiarities,
that were not considered in the somewhat oversimplified
treatment of the section \textbf{\ref{sec6}}.

The first point to explain is the situation with the 1PI
(one-particle-irreducible) graphs.
They were omitted in the treatment of sec. \textbf{\ref{sec6}}
because they do not contribute to the counter-terms $Q$.
This can be easily proved by induction.
Indeed, consider a {\it one-particle-reducible} graph
$$\Gamma=\twopi$$
(external legs are not shown).
Its Feynman amplitude is given
by the product of propagator and two amplitudes for its 1PI parts
$\gamma_1, \gamma_2$.
Assuming by induction hypothesis that $Q$ for all the
one-particle-reducible graphs with smaller number of vertices is
zero, we conclude that the set of box-subgraphs of $\Gamma$,
which can potentially contribute to $Q(\Gamma)$
is just the unification of the corresponding sets for its
1PI parts, plus boxes which surround the whole $\gamma_1$
or $\gamma_2$.
Thus we get:
$$Q(\Gamma)=\Pm\Big\{\left(F(\gamma_1)-Q(\gamma_1)\right)
\left(F(\gamma_2)-Q(\gamma_2)\right)\Big\}\times
\left\{{\begin{picture}(13,9)\put(0,3){\line(1,0){12}}
\end{picture}}\right\}=0$$
because
$$\Pm\Big\{F(\gamma_1)-Q(\gamma_1)\Big\}=
\Pm\Big\{F(\gamma_2)-Q(\gamma_2)\Big\}=0$$
This derivation is clearly based on the fact,
that all divergencies in QFT come from loops,
while the one-particle-reducible graph
is just the product of its components.

The second point to clarify concerns renormalization
of mass and kinetic terms.
To study this subject we need to introduce two more
coupling constants in our treatment in \ref{sec6}
and, correspondingly, two more elementary vertices:
$$\go, \cross$$
the first one corresponds to the mass term
and the second one to kinetic term.
Lagrangian and elementary Feynman vertices look as follows:
$$L(\phi)=\frac{\kappa}{2}(\partial_{\mu}\phi)^2-
\frac{1}{2}m^2\phi^2-\frac{T}{4!}\phi^4$$
$$F\big(\go\big)=1,\ \  F\big(\cross\big)=p^2,\ \
F\left(\oloop\right)=1$$
$$Z\big(\go|T\big)=m^2, \ \ Z\big(\cross|T\big)=\kappa,\ \
Z\left(\oloop|T\right)=-iT $$
Usually one assumes that the physical value
$\tilde\kappa=1$, but its bare value $\kappa$ can
be different.

Now we can write the sum of all correlators which potentially can
contribute to $Q$, i.e. are potentially divergent:
\vspace{-0.5cm}
\be
F(T)=\go+ \cross+ \oloop+ \oneloopmass+ \twoloopmass+\onepi+
\left(\sloop+\ldots\right)+ \left(\sloopm+\ldots\right)+\nn\\
+\left(\siiloop\ +\ldots\right)+\left(\siirloop+\ldots\right)+
\left(\siilloop+\ldots\right)+\ldots \nn
\ee
Since we have two different elementary vertices,
corresponding to two different couplings,
a new external structure should be introduced to separate
contributions to $Q$ into two parts:
$Q$ for graphs with $2$ external legs acquires additional label,
equal to $0$ or $1$.
We have two projectors $\Pm^{(0)}$ and $\Pm^{(1)}$
for the self-energy graphs:
$\Pm^{(0)}$ picks up the singular terms, which are
proportional to $p^0$, while  $\Pm^{(1)}$ picks up terms,
proportional to $p^2$.
Accordingly,
$$Q^{(i)}(\Gamma)=-\Pm^{(i)}\big\{F(\Gamma)+\ldots\big\}$$
Now renormalization process may be done "as usual",
$$\tilde m^2= m^2+ \sum_{\stackrel{\mbox{\tiny{self-energy}}}
{\mbox{\tiny{graphs}}}} Q^{(0)}(\Gamma)$$
$$\tilde \kappa= \kappa+
\sum_{\stackrel{\mbox{\tiny{self-energy}}}
{\mbox{\tiny{graphs}}}} Q^{(1)}(\Gamma)$$
$$-i\tilde T= -iT+\sum_{\stackrel{\mbox{\tiny{graphs with $4$}}}
{\mbox{\tiny{external legs}}}}Q(\Gamma)$$
For more elaborated treatment of renormalization with mass terms,
see refs.\cite{CK,CK2} for $\phi^3$ theory,
ref.\cite{Vass} for $\phi^4$ theory and
refs.\cite{CK2,QED} for QED.

\section{Acknowledgements}
Our work is partly supported  by Dynasty Foundation,
and  the RFBR grant 07-02-00878 (M.S.),
by Federal Nuclear Energy Agency, by RFBR grant 07-02-00645,
by the joint grant 06-01-92059-CE,  by NWO project
047.011.2004.026, by INTAS grant 05-1000008-7865,
by ANR-05-BLAN-0029-01 project (A.M.)
and by the Grant of Support for
the Scientific Schools NSh-8004.2006.2.

\end{document}